# Is artificial consciousness achievable?
# Lessons from the human brain

## Michele Farisco[1,2], Kathinka Evers[1] & Jean-Pierre Changeux[3]

1 Centre for Research Ethics and Bioethics, Department of Public Health and Caring Sciences, Uppsala University, Uppsala, Sweden.
2 Biogem, Biology and Molecular Genetics Institute, Ariano Irpino (AV), Italy
3 Neuroscience Department, Institut Pasteur and Collège de France Paris, France

⋅ Address for Correspondence: Centre for Research Ethics & Bioethics (CRB), Uppsala University, Box 564, SE-751 22 Uppsala. E-Mail: michele.farisco@crb.uu.se

## Abstract

We here analyse the question of developing artificial consciousness from an evolutionary perspective, taking the evolution of the human brain and its relation with consciousness as a reference model or as a benchmark. This kind of analysis reveals several structural and functional features of the human brain that appear to be key for reaching human-like complex conscious experience and that current research on Artificial Intelligence (AI) should take into account in its attempt to develop systems capable of human-like conscious processing. We argue that, even if AI is limited in its ability to emulate human consciousness for both intrinsic (i.e., structural and architectural) and extrinsic (i.e., related to the current stage of scientific and technological knowledge) reasons, taking inspiration from those characteristics of the brain that make human-like conscious processing possible and/or modulate it, is a potentially promising strategy towards developing conscious AI.

Also, it cannot be theoretically excluded that AI research can develop partial or potentially alternative forms of consciousness that are qualitatively different from the human form, and that may be either more or less sophisticated depending on the perspectives. Therefore, we recommend neuroscience-inspired caution in talking about artificial consciousness: since the use of the same word "consciousness" for humans and AI becomes ambiguous and potentially misleading, we propose to clearly specify which level and/or type of consciousness AI research aims to develop, as well as what would be common versus differ in AI conscious processing compared to human conscious experience.

## Introduction

Since Helmholtz, DuBois-Reymond and even Freud pledged the solemn oath (1842) that "no other forces than the common physical chemical ones are active within the organism", there is wide scientific agreement that the brain is a "physico-chemical system" and that "consciousness" is one of its most sophisticated features, even if there is no consensus on explaining how, specifically, this is the case. Therefore, it can be argued, it is theoretically possible that sooner or later one should be able to artificially emulate the brain's functions, including consciousness, through physico-chemical methods. Yet the situation is analogous to the case of "life in the test tube" with the simplest living organisms: all their molecular components are known but up to now nobody has been able to reconstitute a living organism from its dissociated components. The issue is not only theoretical but also, importantly, practical.

The prospect of developing artificial forms of consciousness is increasingly gaining traction as a concrete possibility both in the minds of lay people and of researchers in the field of neuroscience, robotics, AI, neuromorphic computing, philosophy, and their intersection (Blum & Blum, 2023; Butlin et al., 2023; LeDoux et al., 2023; Oliveira, 2022; VanRullen & Kanai, 2021). The challenge of artificial conscious processing raises also social and ethical concerns (Farisco, 2024; Farisco et al., 2023; Hildt, 2023; Metzinger, 2021). Therefore, it is very timely to critically evaluate the feasibility of developing artificial conscious processing from a multidisciplinary perspective, as well as analyzing what that concept might mean. Relevant attempts in this direction have recently been proposed (Aru, Larkum, & Shine, 2023; Godfrey-Smith, 2023; Seth, 2024).

Current discussions about the theoretical conceivability and the technical feasibility of developing artificial conscious processing hinges, to begin with, upon a semantic ambiguity and polysemy of the word "consciousness", including the distinction between phenomenology (i.e., a subjective first-person experience) and underlying physiology (i.e., a third-person access to consciousness)(Evers & Sigman, 2013; Farisco, Laureys, & Evers, 2015; Levine, 1983), and the fundamental distinction between conscious and non-conscious representations (Piccinini, 2022). Also, conscious processing may have different meanings depending on the context of analysis and it has different dimensions, which may possibly exhibit different levels resulting in different profiles of conscious processing (Bayne, Hohwy, & Owen, 2016; Dung & Newen, 2023; Irwin, 2024; Walter, 2021). At the origins, consciousness comes from the Latin *conscientia*, *cum scire*: knowledge in common, oscillating between confidence and connivance, up to the classic "faculty that man has of apprehending his own reality" (Malebranche 1676) or for the neuropsychiatrist Henri Ey "the knowledge of the object by the subject and reciprocally, the reference of the object to the subject itself". Accordingly, the individual is both the subject of his knowledge and the author of it. Lamarck, in 1809, speaks of a singular faculty with which certain animals and even humans are gifted, which he calls "sentiment interieur", approximately inner feeling. More recently Ned Block introduced the distinction between access and phenomenal consciousness. Access consciousness refers to the interaction between different mental states, particularly the availability of one state's content for use in reasoning and rationally guiding capabilities like speech and action; phenomenal consciousness is the subjective feeling of a particular experience, "what it is like to be" in a particular state (Block, 1995). Accordingly, cognition and subjective experience are two central components of conscious processing, which basically may be defined as "sensory awareness of the body, the self, and the world" (Lagercrantz & Changeux, 2009), including "inner, qualitative, subjective states and processes of sentience or awareness" (Searle, 2000). Among the embodied components of conscious processing we may consider, in addition, at the individual level, the ability to express emotions, memory, symbols, language, capacity for autobiographical report and mental time travel, as well as the capacity to introspect and report about one's mental state, and at the social level, sustained inter-individual interactions which give access to various kinds of social relationships such as empathy and sympathy  (Lagercrantz & Changeux, 2009).

Among the many theories and computer science models currently proposed, none of them, in our assessment, reach the overall species-specific aspects of the human higher brain functions (van Rooij et al., 2023). The question arises: can these models reach those aspects with time, when further developed, or is the gap irremediable? In parallel, more and more citizens are confronted with AI simulations of human behaviour, including conscious processing, and feel concerned about it (Lenharo, 2024): the prospect of artificial conscious systems raises the risk of impacting human self-understanding, for instance if AI were to replace humans in performing tasks that require a capacity for awareness. It thus appears necessary to challenge AI models with actual representations of human brain organization and human cognition and behaviour. Therefore, the question is whether or not any theoretical computer science representation of human conscious processing can lead to human-like artificial conscious systems: could machines ever develop a human-like consciousness, or rather a different kind of consciousness, or is it impossible for them to develop consciousness at all? Does the notion of artificial consciousness even make sense, and if so, how? To paraphrase Voltaire: Can a machine awaken?[1]

In the past decades, a large number of models were elaborated mainly by neuroscientists with a more humble aim: to reconstruct elementary functions of the nervous system (e.g., swimming in the leech (Stent et al., 1978) or the lamprey (Grillner et al., 1995)) from known anatomical and physiological building blocks. Some of these models have even been designed to simulate more elaborated cognitive tasks like the Wisconsin Card sorting task (Dehaene & Changeux, 2011) and even trace vs. delay conditioning (Grover et al., 2022). It is necessary to further develop the interface between AI, philosophy and neuroscience, which thus far has resulted in a mutual epistemic and methodological enrichment (Alexandre et al., 2020; Farisco et al., 2023; Floreano, Ijspeert, & Schaal, 2014; Floreano & Mattiussi, 2008; Hassabis, Kumaran, Summerfield, & Botvinick, 2017; Momennejad, 2023; Poo, 2018; Zador et al., 2023). In fact, although significant, this collaboration is still insufficient to address the issue of artificial consciousness. The crucial, still open question is: what kind of concrete similarities vs. differences between AI and the brain may need to be examined and accounted for to more adequately approach artificial conscious processing? In other words, what is the right 'level of description' to either model, or even generate artificial conscious processing given what we know about conscious processing in the human brain?

Moreover, in the neuroscience field, the word "consciousness" remains rather ill-defined and, as we shall see below, human conscious processing is not an all-or-none irreducible feature but one that develops stepwise (Changeux, 2006, 2017; Lagercrantz & Changeux, 2009; Tomasello, 2022; Verschure, 2016). Given these different possible developmental stages, AI attempts to develop artificial conscious processing should precisely specify which one (if any) of these developmental stages is selected.

In this paper we want to re-evaluate the issue of artificial consciousness within the context of our present knowledge of the biological brain, taking a pragmatic approach to the conceivability and feasibility of developing artificial consciousness and using the human brain as a reference model or benchmark. We aim to complement recent attempts in this direction (Aru et al., 2023; Godfrey-Smith, 2023) with a more encompassing analysis of the biological multilevel complexity of the human brain in relation to its evolution, not only in order to progress in the understanding of conscious processing itself but also to eventually inspire ongoing AI research aimed at developing artificial conscious processing. Accordingly, our aim is theoretical and philosophical but also highly practical as an engineering issue: we review scientific evidence about some features of the brain that are key in enabling human consciousness or in modulating it (or both), and we argue for the utility of taking inspiration from these features for advancing towards the development of conscious AI systems.

---

[1] The philosophers of the Enlightenment already wondered: what in the brain's architecture might explain why and how it became conscious? What made matter awaken? Cf. e.g., a letter from Voltaire to d'Alembert, November 28, 1762.

We do not claim that it is necessary to integrate the mechanisms identified for conscious processing in the human brain to develop artificial consciousness. In fact, we recognize that artificial features of conscious processing that are different from the brain ones cannot theoretically be excluded offhand. What we propose is rather to take the presently identified brain mechanisms of conscious processing as a benchmark in order to pragmatically advance in the building up of artificial models able to simulate accessible features of conscious processing in humans. Given the high controversy around the possibility to build up an artificial consciousness unrelated to brain mechanisms and the related risk of ending up in overly abstract views that are not sufficiently informed by empirical data, we think that starting from the biology of consciousness is a more productive strategy.

A question we may nevertheless ask is what are the benefits to pursue artificial consciousness in the first place, for science, or society at large? There are different possible answers. On an epistemological level, consistently with the medieval scholastic view reiterated by i.a. Paul Valéry that "we can actually understand only what we can build", it is clear that to elaborate artificial models of some concrete features of conscious processing could perhaps eventually allow us to better understand biological consciousness in general, whether in terms of similarities or differences. At a technical level, it is possible that the development of artificial consciousness would be a game-changer in AI, for instance giving AI the capacity for intentionality and theory of mind, and for anticipating the consequences of its own "actions". At the societal and ethical level, especially the last points could arguably help AI to better inform humans about potential negative impacts on society, and to help avoid them while favouring positive impacts. Of course, on the negative side, intentionality in machines might not at all favour human interests any more than human intentionality has favoured out-group individuals or species, or indeed the planet as a whole. This is indeed a discussion that would merit deeper analyses, but it is beyond the aim of the present paper.

In the following, we will summarize relevant evolutionary, structural, and functional properties of the human brain that are of specific relevance to this discussion (for a recent overview, see (Barron, Halina, & Klein, 2023). Against that background, we will outline what the brain may inspire to current research on AI for advancing towards artificial conscious systems.

Finally, concerning the conceivability and feasibility of developing artificial consciousness, we will distinguish between:

(a) the replicability of human consciousness (which we exclude, at least in the present state of AI-development, a stance which is scarcely controversial);

(b) the possibility of developing an artificial conscious processing that may bear some resemblances but still is profoundly different than human (which we do not exclude in principle, but consider difficult to elaborate for both conceptual and empirical reasons).

In the end, this paper starts from a selective examination of data from brain sciences with the aim to propose an approach to AI consciousness alternative to what appears to be the leading one today. This approach may be qualified as theory-based because it relies not upon experimental data but on selected components of *a priori* scientific theories which are then applied to AI systems (Butlin et al., 2023). Our approach, on the opposite, consists in starting from empirically established brain mechanisms and processes which are directly relevant to human consciousness and infer from them hardware building blocks or algorithms that are relevant and perhaps even necessary (if not sufficient) to the development of artificial conscious processing.

## 1. Extents and ways in which AI has been inspired by understanding of the brain

### 1.1 Computational models

Presocratic Greek philosophers already stated that any description of reality is produced by human beings (our brain) through models which necessarily displays physical limits (Changeux & Connes, 1995). This is also a logical limitation. As Kant (1781) argued, all our experiences make essential reference to our own, perforce finite, perspectives that we can never transcend, which means that we are, in a manner of speaking, prisoners of our brains (Evers, 2009). In other words, we are epistemically limited because of our finitude and the physical constraints of our brains to produce models. Accordingly any mathematical modeling, including AI that relies on and is informed by computational models, shall never be able to give an "exhaustive" description of reality, physical or biological. The issue is thus whether and to what extent any model that assumes that a brain function/feature like conscious processing may be implemented in exactly the same way in different physical structures, either biological or artificial (i.e., functionalism), can be useful to partially or fully describe or simulate the brain (e.g., generating testable hypotheses about human consciousness, such as, for example, the Global Neuronal Workspace theory and its experimental evaluation (Mashour, P. Roelfsema, J.-P. Changeux, & S. Dehaene, 2020b)), notwithstanding the fact that, even if potentially useful (Smaldino, 2017), any biological model of the brain today is an oversimplification of neuroscientific data and of their actual biological complexity (Chirimuuta, 2024). This is not to deny that brain models may be useful and adequate even if limited in the number of details they contain, depending on the specific aspects of the brain that are modelled and on the relevant level of description. It is theoretically possible, for instance, that not all the lower-level details are necessary in order to reproduce, predict, or simulate some higher level properties, and therefore that higher levels of description of the system provide more relevant and more sufficient information (Hoel, 2017) (Rosas et al., 2020). Yet, if the goal is a comprehensive description or even a simulation of the whole brain, then any computational model would be insufficient (Farisco, Kotaleski, & Evers, 2018).

The relevance of low-level neural organisation for the simulation of conscious processing has been denied by functionalist philosophers (Butlin et al., 2023). Recently, Peter Godfrey-Smith has argued that the functional similarity of two systems is a matter of degree (i.e., it depends on the extent that a system needs to be understood, in coarser or finer-grained ways) (Godfrey-Smith, 2023). The crucial point is what a degree of similarity is necessary for duplicating an entity like conscious processing. Multiple realizability is the thesis that the same kinds of mental capacities can be manifested by systems with different physical architectures (Cao, 2022). This thesis has been contested. For instance, following Ned Block (Block, 1997), Rosa Cao recently argued that strict functionalism imposes quite stringent constraints on the underlying physical structure rather than eventually allowing multiple realizability. In fact, complex integrated functions (like consciousness) impose more constraints, including at fine-grained levels, than functions that can be decomposed into simpler independent functions (Cao, 2022).

Other theoretical accounts of consciousness have a somehow ambiguous critical stance towards functionalism and multiple realizability. This is the case, for instance, of the Integrated Information Theory (IIT) (Albantakis et al., 2023). IIT relates conscious processing to "integrated information" (i.e., the amount of information generated by a complex of elements, over and above the information generated by its parts). Intrinsic information is defined as what makes differences within a system. Conscious processing is ultimately identical with intrinsic information: a system is conscious if it generates information over and above its constituting parts and independently from external observers-interpreters. This is the reason why, according to IIT, "a digital simulation of the brain cannot be conscious", neither in principle or in practice. On the other hand, a neuromorphic silicon-made computer could be conscious, because it could be composed of neuron-like elements intrinsically existing and characterized by conceptual structures (i.e., cause-effect repertoires) similar to ours (Tononi, 2015).

Therefore, IIT is against functionalism, arguing, in the spirit of Edelman (Edelman, 1992; Tononi & Edelman, 1998), that an exclusive focus on functions ignoring the physical structure cannot explain consciousness (Tononi, 2015). Particularly, re-entry processes are crucial for explaining consciousness: only systems with feedback loops can integrate information, while feed-forward systems cannot become conscious. Thus IIT is not functionalist because it stresses the crucial role of physical components ~~substrate~~ necessary for information generation and integration, that is, for conscious processing. Furthermore, according to IIT, a system that functions like a conscious human is conscious only if it uses the same architecture (i.e., re-entrant) as humans. Even if not functionalist, IIT eventually admits the possibility of replicating consciousness in different systems.

## 1.2 Artificial Neural Networks

Historically, from the early times of AI, Pitts and McCulloch referred to networks of idealized artificial neurons (McCulloch & Pitts, 1943). The connectionist program was subsequently negatively affected by Marvin Minsky's seminal critique (Minsky & Papert, 2017) to then gradually come back until the explosion in use and popularity of Artificial Neural Networks (ANNs) in recent times (LeCun, Bengio, & Hinton, 2015). Decades of computer research developed on this basis moving from simplistic to complex architecture, from a single to multiple layers of artificial neurons – from the perceptron up to deep learning (LeCun et al., 2015) and the billions of parameters of Large Language Models (LMMs) (e.g., ChatGTP). Not only the symbolic approach ("Good Old-Fashioned AI" or GOFAI, including the "Logic Theorist" created by Newell and Simon), which was prevalent at the beginning of AI research and that aimed to reproduce the logical aspects of intelligence at a high functional level while neglecting the underlying brain mechanisms, but ultimately also the ANNs program does not fully account for the complexity of brain architecture (Moulin-Frier et al., 2016), if any reference to it is included.

In fact, the notion of neuron referred to in ANNs (i.e., a mathematical function which models biological neurons) is much simpler than its biological counterpart. For instance, basal and apical dendritic compartments of cortical pyramidal neurons connect to feedforward and feedback information processing respectively (Aru, Suzuki, & Larkum, 2020). This complexity intrinsic to biological neurons is partly taken into account in most modern ANNs, even if structured neurons for AI applications is an active field of research (Haider et al., 2021; Max et al., 2023; Senn et al., 2023). Other characteristics of the brain, like the role of GABAergic interneurons and the modulation by neurotransmitters like acetylcholine and dopamine (Changeux & Lou, 2011), or the capability of pyramidal cells to have two functionally distinct sets of inputs (one about which the neuron transmits information, and one that can selectively amplify that transmission when it is useful to do so in the context of information being transmitted by other neurons) (Phillips, 2023) are translated into large scale brain models like functional patterns (Eliasmith et al., 2012; Humphries, Khamassi, & Gurney, 2012). As mentioned above with reference to the multiple realizability thesis, the question is whether this functional emulation captures the right elements, processes and properties of the world to produce an empirically adequate description of the target object (e.g., the brain) and on this basis to possess its same properties (e.g., the same causal effect of the neuronal underpinnings of human features, like conscious processing). In other words, the question is whether it is possible to reproduce the features of real world objects, like the brain, in computing systems, including those built on digital electronic principles, different from the analog biological principles of the brain.

The recent developments of AI (e.g., ChatGPT, Sora, Dall-E, Stable Diffusion) illustrate further the extent to which such a basically algorithmic approach can be successful, even if the discussion is open about its limitation (Mitchell, 2023; Mitchell & Krakauer, 2023; Shanahan, 2024; Shanahan, Crosby, Beyret, & Cheke, 2021). Similarly, for years, computational "functionalist" descriptions of cognitive processes deliberately excluding any reference to the human brain were abundantly produced. Again, this connects to the question about multiple realizability of cognitive processes: can the same outcome be achieved with completely different biological, cognitive, or computational mechanisms? (Melis & Raihani, 2023)). Are we facing analogy rather than homology? The challenge is whether a *formal algorithmic based computational* approach, up to now successful in a number of applications, in particular the so-called neuromorphic hardware (Petrovici et al., 2014; Poo, 2018), will ever be sufficient to approach human-like conscious processing, an alternative form of conscious processing, or eventually no consciousness at all (Kleiner, 2024).

## 2. Embodiment of conscious processing: hierarchy and parallelism of nested levels of organization

The hardware of the most common computers is made up of microprocessors, which are fabricated on semiconductor transistors (including oxide-based memristors, spintronic memories, and threshold switches) or neuromorphic substrates (Billaudelle et al., 2020; Pfeil et al., 2013) integrated into the brain circuit chips. These physical elementary components of the hardware are made up of a few chemical elements and compute at a high speed, much higher than the brain (see below), despite the fact that information transfer between two logical components in a system requires multiple stages of encoding, refreshing, and decoding, in addition to the pure velocity of electrical signals (which is about half the speed of light in standard circuit boards). In contrast, the brain is built from multiple nested levels of organization from highly diverse chemical elementary components, much more diverse than in any computer hardware. It is true that computational models like neuromorphic systems may be designed to account for such multiple nested levels of brain's organization (Boybat et al., 2018; Sandved-Smith et al., 2021; Wang, Agrawal, Yu, & Roy, 2021), but the fact remains that their primary outcome is a limited emulation of brain functions while these computational models cannot account for the biochemical diversity of the human brain and its pharmacology, including the diseases which possibly affect conscious states and conscious processing (Raiteri, 2006). In fact, it is possible that the high level of biochemical diversity characteristic of the brain plays a role both in making possible and in modulating conscious processing (e.g., with psychotropic drugs). This possible role of the brain's biochemical diversity for enabling conscious processing should be further explored in order to eventually identify relevant architectural and functional principles to be translated into the development of conscious AI systems.

In the real brain, the molecular level and its constraint play a critical role often insufficiently recognized. Proteins are the critical components. These macromolecules are made up of strings of amino acids which fold in a highly sophisticated organization able to create specific binding sites for a broad diversity of ligand including metabolites, neurotransmitters, lipids, and DNA. Among them there are enzymes which catalyse/degrade key reactions of the cell metabolism, the cytoskeleton and motor elements, DNA binding proteins like transcription factors ion channels, and, most of all, neurotransmitter receptors together with the many pharmacological agents which interact with them. The number of different proteins in all organisms on earth is estimated to be $10^{10}$–$10^{12}$. In the human brain 3.710-3.933 proteins are involved in cognitive trajectories (Wingo et al., 2019) which is a significant fraction of the total number of genes, that approximates 20.000 in humans. There is a fundamental heterogeneity of the protein components of the organism which contribute to brain metabolism and regulation. This is true also for the control of the states of consciousness and their diversity, as assessed by the global chemical modulation of sleep and wakefulness or the diversity of drugs that cause altered states of consciousness (Jékely, 2021). In all cases the modulation ultimately takes place at the level of proteins. Access to conscious processing in the human brain thus requires a rich and dedicated biochemical context that is widely absent from the current AI attempts to produce artificial conscious processing: computers have no pharmacology and no neuro-psychiatric diseases. Data indicate a crucial significance of the molecular components in the understanding of life and thus of biological conscious processing, as increasingly recognized in both neuroscience and philosophy (Aru et al., 2023; Godfrey-Smith, 2023; Seth, 2021, 2024; Thompson, 2018). Indeed, in principle the crucial role played by pharmacological factors for human consciousness does not oppose the fact that AI systems may instantiate alternative, non-human like forms of consciousness (Arsiwalla et al., 2023). Our reference to the abovementioned molecular and pharmacological conditions is to identify what factors current AI systems lack that may be taken as inspiration for advancing towards artificial consciousness.

Another factor that is related to the different physical architectures of biological brains and computers and that presently makes a difference between them is the velocity of information processing. As mentioned above, the speed of information processing in the brain is lower than the speed of sound (~ 343 m/s), while computers work at a speed which approaches the speed of light ($3x10^{10}$ m/s). This creates an insurmountable difference in the speed of propagation of information in computers vs. brains. The conduction velocity of the nerve impulse may reach 120 m/s (depending on the diameter of the axon) and is largely consequence of the allosteric transition of the ion channels in some myelinated axons. But additional constraints are imposed by the synapse which creates delays in the millisecond range including the local diffusion of the neurotransmitter and the allosteric transition of the receptor activation process (Burke et al., 2024). Because of the cumulation of several of these successive delays, psychological timescales are in the order of 100 (50-200 ms) (Grondin, 2001). Last, access of a peripheral sensory signal to consciousness may take delays in the order of 300ms (Dehaene & Changeux, 2011). In short, standard computers process information up to 8 orders of magnitude faster than the human brain. It is expected that for highly standardized tasks - such as playing chess – the computer is more efficient. This is in part due to different computational strategies, with the computer being able to test millions of possibilities in parallel while the human brain is able to test only a few during psychological time scales. As a consequence, the computers are able to process amounts of data so huge that they are not accessible to the brain in a human life time.

Moreover, despite some significant improvements obtained through neuromorphic systems, including increased computational power per unit of energy consumed (Cramer et al., 2022; Esser et al., 2016; Göltz, Kriener, Sabado, & Petrovici, 2021; Park, Lee, & Jeon, 2019), the human brain consumes orders of magnitude less energy than standard computers, even more in the case of massive models like GPT-4. The brain is not only much more energy efficient, but also much more sample efficient: deep-learning models are considered sample-inefficient when compared to human learning due to the huge amounts of training data they require to achieve human-level performance in a task (Botvinick et al., 2019; Botvinick, Wang, Dabney, Miller, & Kurth-Nelson, 2020; Waldrop, 2019).

Thus, the exceptional success of AI in simulating brain processes results largely from the speed of processing of increasingly sophisticated algorithms and computer programs. This huge difference in processing speed may be interpreted as an argument against artificial consciousness (Pennartz, 2024): since AI processes much more data in a much faster way than human brains, it may be that the parallel processing of data is sufficient for AI to accomplish those tasks that require integrated (i.e., conscious) processes in humans. In fact, one of the defining features of conscious processing is the integration of the information processed in different neuronal populations, which eventually result in a unified multimodal experience. This integration can be fundamentally defined as a broadcasting of representations mediated by an "ignition" pattern of activity, i.e. of organized neuronal activity patterns (Mashour et al., 2020b). From an evolutionary point of view, because of intrinsic biological constraints that determine the quantity and complexity of the sensory modalities received, as well as the timescale of operation of the cognitive and control systems, this integrative and broadcasting activity serves the need of handling complex information, that is information composed by multiple elements that cannot be processed in parallel by the brain. In the case of AI, the capacity of processing complex information produces integration and broadcasting that is not at the same level and in the same modality as in the human brain. In fact, modern AI systems are hugely modular, and they are based on the broadcasting of representations, but the constraints that characterize AI modules and eventually the capacity of the whole AI system to compute in parallel are significantly different than those characterizing the human brain.

This point may be illustrated in terms of quality (i.e., the kind of operations that the modules are capable of performing) and in terms of quantity (i.e., the amount of data that the modules are capable of processing). From a qualitative point of view, the diversity of the modules composing an AI system may consist in the specific task they perform and may approximate the diversity of sensory modalities the brain has access to, while those modules cannot rely on the different valence attributed to the processed information and to its representation, like the human brain can do. This is why, for instance, deep convolutional networks are eventually collections of "abstract concepts that lack experiential content, imagination and integral perception of the environment" (Pennartz, 2024). Therefore, at least to date, compared to the human brain AI modules are different in nature, because of their inability to attribute experiential value to information. To use the abovementioned differentiation between access and phenomenal consciousness, modular AI systems may reach the capacity to broadcast information among different modules (i.e., they may approximate access consciousness), while they appear presently unable to replicate phenomenal conscious processing or subjective experience. Dehaene, Lau, and Kouider propose a similar interpretation (Dehaene, Lau, & Kouider, 2017). They conclude that current AI systems may have the capacity for global availability of information (i.e., the ability to select, access, and report information) while they lack the higher level capacity for metacognition (i.e., the ability for self-monitoring and confidence estimation) (Lou, Changeux, & Rosenstand, 2016).

This interpretation converges with the distinction between the implementation of attention and access consciousness in AI and the (current) impossibility to engineer phenomenal consciousness. Along the same line, (Montemayor, 2023) outlines that present AI systems are limited to representational values that are instrumental to satisfy epistemic needs, while current AI has no experience of moral and aesthetic needs which are based on the intrinsic value of phenomenal consciousness and subjective awareness..

From a quantitative point of view, the capacity for memory storage and the computing power of AI systems is so much greater than that of the human brain that it eventually makes the system capable of performing the same tasks which in humans require a conscious effort at a completely non-conscious level. In fact, conscious processing may be considered an evolutionary strategy to enhance the survival of biological systems with limited capacities for memory storage and data processing.

In conclusion, different qualitative and quantitative constraints result in different operations (conscious vs non-conscious). Interestingly, when applying the 'right kind of constraints to an artificial model', some properties can develop as a result (Ali & al., 2022). It may be that conscious processing derives as a necessity from the constrained computational capacities of the human brain.

An additional question is what still makes the brain highly performing in psychological time scales up to conscious processing. The answer is mainly in the hardware architecture, which in turn may impact the kind of computations that biological brains are capable of. In the brain, the neurons and their neurites as well as their sophisticated connectivity are far more complex than the idealized artificial neurons. The biological neurons originate from definite supramolecular assemblies which create an important chemical complexity and diversity of cell types (maybe up to a thousand), the neuron types differing also by their shape, their axonal-dendritic arborizations and the connections they establish. An important chemical difference are the neurotransmitters (up to hundreds) they synthesize and release (Jékely, 2021).

Depending on the neurotransmitter they release, two main classes of neurons may be distinguished (i.e., excitatory vs. inhibitory). To pass a simple-minded conscious task, such as trace vs delay conditioning mentioned above, several superposed levels of organization are required and also the contribution of both inhibitory and excitatory neurons appears necessary (Grover et al., 2022; Volzhenin, Changeux, & Dumas, 2022). It is true that also in the case of deep network the weight matrix contains as many negative (inhibitory) entries than positive (excitatory) ones, and that there are neurobiologically-plausible backpropagation algorithms for deep neural networks that include both excitatory and inhibitory neurons (Guerguiev, Lillicrap, & Richards, 2017; Lillicrap, Santoro, Marris, Akerman, & Hinton, 2020; Scellier & Bengio, 2017; Whittington & Bogacz, 2017), but, as mentioned above, the diversity of the types of biological neurons is so vast that we still do not completely account for it, so those computational algorithms may eventually miss architectural and functional details of the brain that are crucial for conscious processing.

Inhibitory and excitatory neurons establish basic neuronal networks levels, which themselves establish higher levels of nested networks of neurons, and so on. Such sophisticated connectomic relationships might contribute to the genesis of cognitive functions and even consciousness. For instance a cellular theory of consciousness has been suggested (Aru et al., 2020) based upon the dendritic integration of bottom-up and top-down data streams originating from thalamo-cortical neuronal circuits. Such hardware complexity is most often absent from the most elaborated super-computers, even if exceptions do exist, notably in neuromorphic hardware making use of cortical microcircuits (Haider et al., 2021;

Max et al., 2023; Senn et al., 2023). Also, it is true that systems like LLMs show emergent abilities likely deriving from emergent structures that are present in their learned weights. In fact, at an architectural level, they are arranged as a large number of layers, which potentially allows for the sorts of organization (i.e., hierarchical, nested, and multilevel) that arguably plays a crucial role for conscious processing in the brain. Therefore, we cannot exclude in principle that large AI systems have the potentiality to develop some forms of conscious processing with characteristics shaped by the specific components of the AI systems' structural and functional organization.

In conclusion, the hierarchical, nested, and multilevel organization of the brain indicates some architectural and functional features that current AI only partially emulates and that we propose should be further clarified and engineered in order to advance towards the development of conscious AI, including massive biochemical and neuronal diversities (i.e., excitatory vs inhibitory neurons, connectomic relationship, circuits, thalamo-cortical loops), and a variety of bottom-up/top-down regulations between levels. Among other things, these organizational features constrain the speed of information processing in the brain, up to several orders of magnitudes slower than in AI: this difference may eventually make conscious processing not necessary for AI to execute those tasks that require conscious efforts in humans.

## 3. Evolution: from brain architecture to culture

### 3.1 Genetic basis and epigenetic development of the brain

Human-made computers are made up serially from prewired components rigidly associated into substantially time invariant machines with identical hardware. On the other hand, the brain's "hardware" developed in the course of Darwinian evolution from sponges to the human brain following multiple steps of selection which led to a progressively increased complexity of the architecture and thus of behavioral abilities (Kelty-Stephen et al., 2022). The selection took place along essentially different environment conditions, including physical, social and possibly cultural factors. As a consequence, new levels became nested upon former ones without necessary rational (i.e., predictable) relationships. The architecture of the brain thus recapitulates the multiple incidents of its evolutionary history, and it is not the outcome of a rational and basic human design like AI systems, which may be viewed as the end products of a human directed cultural evolution (possibly including variation and selection processes) (Mesoudi et al., 2013; Solée et al., 2013; Valverde, 2016).

Also, a consequence of the brain evolution at the genomic level is that if each novel acquisition builds upon a former one a strong conservation of the genome might take place through evolution: to illustrate, the drosophila genome is 60% homologous to that of humans, and about 75% of the genes responsible for human diseases have homologs in flies (Ugur, Chen, & Bellen, 2016). Within this context the hominization of the brain took place on the basis of a few genetic events (still mostly unidentified) without a dramatic reorganization of the genome. Yet these changes have been sufficient for the broadcasting of information among distant thalamocortical regions (i.e., the global neuronal workspace (GNW)) to arise and consciousness to acquire the specific features of the adult human brain (Changeux, Goulas, & Hilgetag, 2021; Hublin & Changeux, 2022).

To escape from the conservative genetic entanglement that frames Darwinian evolution in paleontological time-scale (million years), a new kind of much faster (days, hours) non genetic evolution developed. It takes place during postnatal development (ca 25 years in humans), where about half of the total number of synaptic connections are formed and contribute to the formation and shaping of the connectomic "hardware" of the adult human brain. The CCD theory (Changeux, Courrège, & Danchin, 1973), which formally accounts for this epigenetic evolution, relies upon the variability of developing inter-neuronal connections and selection-elimination processes, which formally resemble an evolutionary "Darwinian" process by variation-selection (Changeux & Danchin, 1976; Edelman & Mountcastle, 1978; Kasthuri & Lichtman, 2003). The development of the brain - from newborns to adults – then progresses as a multistep nested foliation resulting from successive waves of synapse outgrowth and selection. This becomes a mechanism to store a gigantic number of long-term memories resulting from the interactions between the brain and its physical, social and cultural environment in the course of postnatal development. The model accounts for a property of the human brain that has enormous social relevance (Evers, 2015) but that has largely been under-evaluated especially in neuroscience: the epigenetic variability of the adult brain's connectivity. According to the CCD variability theorem, "different learning inputs may produce different connective organizations and neuronal functioning abilities, but the same behavioral abilities". Thus, the neuronal connectivity code exhibits "degeneracy" (Edelman & Gally, 2001; Edelman & Mountcastle, 1978; G. Tononi, Sporns, & Edelman, 1999), that is, different code words (connection patterns) may carry the same meaning (function). Such variability together with that of the multiple diverse experiences encountered in the course of development superimposes on that created by the variability of the genome. Ultimately, the history of every human individual becomes incorporated into the hardware of the brain. This essential variability, taking place even between brains of isogenic individuals, strikingly differs from the repeatability of industrial computer hardwares, even if a limited form of variability may occur as AI systems are trained on specific tasks as well as in partly analog neuromorphic computers.

The extension of the postnatal period of development in humans has important consequences in the framework of the theory of synapse selection since it leads to the acquisition and transmission of individual experience and the genesis and internalization of culture in a social environment. The acquisition of reading and writing may be viewed as a typical example of epigenetically developed "cultural circuits" in the brain's hardware. Comparative behavioral and brain imaging studies in illiterate vs. literate subjects reveal striking differences in brain functional connectomics (Castro-Caldas, Petersson, Reis, Stone-Elander, & Ingvar, 1998; Dehaene et al., 2010). Cultural abilities are necessarily and spontaneously learned at each generation from adults to children and epigenetically transmitted from generation to generation, starting even in the womb of the mother, until the adult stage. This spontaneous genesis of cultures does not (yet) make sense with present days computers, even if Large Language Models (LLMs) may implement similar dynamics ( https://arxiv.org/pdf/2402.11271.pdf). In fact, AI (e.g., LLMs) may be pre-trained on vast amounts of data, which may be interpreted as a form of cultural acquisition. Yet there is a crucial difference with the human cultural acquisition, which cannot be reduced to a passive reception of data, but it is a creative re-elaboration of the data received between individuals and from one generation to the other. These dynamics depend on the fact that the brain is not a passive repository of information: the same information may have a different impact on different brains or on the same brain in different times, because perception is modulated by the brain's spontaneous activity, which is more than noise in the internal functioning of the system (see below). As a major consequence of cultural evolution, the acquisition of skills and practices associated with the symbolic experience and emotional labeling characteristic of rational thinking (science), rules of conduct (ethics), and shared feelings (art) become stably internalized epigenetically in the connectivity of the brain and become exclusive of others. Differences in cultures take place between populations with important consequences in the history of the human species. They contribute to what Bourdieu refers to as the 'habitus' of the subject: internalized routine modes of perception, action, and evaluation, which often last a lifetime (Changeux, 2006). This learned regularity of sensorimotor sequential patterns progressively acquired through embodied procedural dynamics is lacking in symbolic systems and in LLMs, while it is currently researched in Embodied AI/Robotics (Gupta, Savarese, Ganguli, & Fei-Fei, 2021). Furthermore, networks that learn to control a robotic agent have been designed to follow an evolutionary process of "synaptic" growth, variation and selection (Miglino, Lund, & Nolfi, 1995; Nolfi, Parisi, & Elman, 1994). Even more so, hyperparameter adaptation is standard in modern AI including changing the network structure, number and nature of layers and their feedback loops, sparsity, activation functions, receptive fields, etc. Evolutionary methods have recently become explicitly part of the repertoire of modern modeling research (Jordan, Schmidt, Senn, & Petrovici, 2021) (Jordan et al 2021).

In conclusion, despite the recent progress towards the implementation of evolutionary approaches in a number of AI technologies as summarized above, today AI still lacks an evolutionary development strictly homologous to that of the human brain.

## 3.2 AI and evolution: consequences for artificial consciousness

Is the lack of an evolution homologous to the human a stumbling block for AI to achieve a human-like conscious processing? Even if AI lacks a strictly phylogenetic evolution (being human made), it may have an ontogenetic development (e.g., through unsupervised learning) that in principle can make possible for AI to achieve capacities not originally programmed by the developers. This would be the case, for instance, of deep reinforcement learning systems (Silver et al., 2018; Vinyals et al., 2019).

Can we say that this form of AI is able to learn from experience, and so to develop an artificial form of "habitus", "temperament", and "character"? In fact, deep reinforcement learning systems includes various kinds of feedbacks and error signals from processed data, which combined with short- and long-term memory capacities may be interpreted as a form of experience.

What about epigenetic and cultural evolution? Even though there are recent computational studies on cultural evolution that show how AI agents can share information from each other and learn to perform tasks (arXiv:2206.05060) (Colas, Karch, Moulin-Frier, & Oudeyer, 2022), at this stage AI mostly lacks an epigenetic development homologous to that of humans. Since human experience and culture derive from epigenesis, it basically follows that AI (still) lacks experience and culture, at least of the same kind and with the same sophistication as in humans.

The evolutionary analysis (Kanaev, 2022) together with the examination of the developmental data for the human newborn (Lagercrantz & Changeux, 2009, 2010) have suggested that human "conscious processing" develops stepwise (Changeux, 2006, 2017).

i) A lowest level of 'minimal consciousness' for simple organisms, like rats or mice, is characterized by the capacity to display spontaneous motor activity and to create representations, for instance, from visual and auditory experience, to store them in long-term memory and use them, for instance, for approach and avoidance behaviour and for what is referred to as exploratory behaviour.

The 25–30-week preterm fetus, according to (Lagercrantz, 2016), would have reached a stage of brain maturation analogous (though not identical) to a newborn rat/mouse; he/she may process tactile and painful stimuli in the sensory cortex (Bartocci, Bergqvist, Lagercrantz, & Anand, 2006), would perceive pain and thus would show signs of minimal consciousness;

(ii) 'recursive consciousness' (Zelazo, Craik, & Booth, 2004), which is present, for instance, in vervet monkeys (possibly also in some birds), manifests itself by functional use of objects and by proto-declarative pointing; organisms at this level may display elaborate social interactions, imitation, social referencing and joint attention; they possess the capacity to hold several mental representations in memory simultaneously, and are able to evaluate relations of self; they display elementary forms of recursivity in the handling of representations, yet without mutual understanding; along these lines, the newborn infant exhibits sensory awareness, ability to express emotions and processes mental representations (i.e., of a pacifier); he/she would already differentiate between self- and non-self touch (Rochat, 2003).

(iii) explicit 'self-consciousness' develops in infants at the end of the second year, together with working and episodic memory and some basic aspects of language; it is characterized by self-recognition in mirror tests and by the use of single arbitrary rules with self–other distinction (Lou, Changeux, & Rosenstand, 2017; Posner & Rothbart, 2007); to some extent chimpanzees might reach this level (Boakes, 1984);

(iv) 'reflective consciousness', theory of mind and full conscious experience, with first person ontology and reportability, reaches full development in humans and develops following 3–5 years in children.

At birth, all major long-distance fiber tracts are already in place (Dubois, Kostovic, & Judas, 2015), although still immature. An electrophysiological signature of conscious processing— homologous to GNW ignition in adult humans—was recorded in 5-, 12-, and 15-month-old babies (Dehaene-Lambertz & Spelke, 2015; Kouider et al., 2013).

In short, subjective experience arguably derives from the temporally extended process of epigenetic development (i.e., a change in the brain connectomics which eventually alter the number of neurons and their connections in the brain network). In other words, consciousness is a process, which depends on and is shaped by the (more or less) temporally extended interactions between the organism (particularly its brain) and the environment.

Present AI has a limited form of this kind of interactive development, which means that social and emotional significance and valence of information are largely missing from or only limitedly present in AI. In fact, AI does not experience the world, nor does it have a theory of mind (Pennartz, 2024): digital AI processes information in a format different from humans, which makes the world that AI can access significantly different from the world that humans have access to. There is a lot of research on analogue computing and AI which in principle may allow it to develop a more human-like experience, but results are still preliminary.

Moreover, given these different possible developmental stages of consciousness, attempts to develop artificial conscious processing should specify which particular stage they target.

Up to now, the diverse AIs including the neuromorphic approaches are not expected to display all the traits of full conscious experience in humans. They may nevertheless qualify for some characteristic features such as a sustained working memory trace, global integration and decision making. They may reach the level of what we have referred to as 'minimal consciousness', or possibly some features of 'recursive consciousness', but neither 'self consciousness' or 'reflective consciousness' appears implementable in artificial systems at present.

Alternatively, the same hypothetical conscious AI may be considered as subject to progressive development. Although it is not the main trend in AI, Alan Turing already proposed in 1950 to consider development as a paradigm for creating and teaching an AI as we could teach a child.

In conclusion, current AI systems, specifically neuro-mimetic robotic systems simulating specific brain's computational models, are limited in producing variability from the same functional organization of a computational model resulting in darwinian biological evolution and epigenetic development through selection and amplification. Also, current AI systems lack the brain's long postnatal development leading to multiple nested epigenetic synapse selections as "inprints" from the physical, biological and socio-cultural environment. As a consequence of the darwinian mechanism of synaptic selection characteristic of the brain, different connection patterns may carry the same function, leading to plasticity, individual differences and creativity.

This means that to date AI has a limited capacity for an evolutionary and epigenetic development analogous to the human brain. We cannot on that basis alone theoretically exclude that deep reinforcement learning systems may one day come to respond to learning and experience in a manner that can be described as an epigenetic relationship with the environment. Yet this may be a limitation also for a theoretical computational approach aiming at extracting the underlying principles of conscious processing for replicating them in artificial systems (Blum & Blum, 2023).

## 4. Spontaneous activity and creativity

To date computers (and AI in general) operate prevalently, even if not exclusively, in an input-output mode. This is strikingly not the case for the human brain which works in a projective – or predictive - mode constantly testing hypotheses (or pre-representations) on the world including itself (Changeux, 1986; Friston et al., 2016; Pezzulo, Parr, Cisek, Clark, & Friston, 2024). This projective/predictive mode relies on the fact that the brain possesses an intrinsic - spontaneous – activity (Dehaene & Changeux, 2005) (i.e., a baseline activity independent from external stimulation). The earliest forms of animals do exhibit a nervous system where spontaneously active neurons can be recorded (e.g., jelly fish, hydra). Such spontaneous oscillators (or pacemakers) result from universal molecular circuits consisting of two fluctuating channels present throughout the animal world up to the human brain. There it was demonstrated early on by EEG (Berger, 1929) and more recently by other brain recordings the existence of a "resting state" or "default" activity distributed in a set of regions that involve mostly association cortex and paralimbic regions (Buckner & Krienen, 2013; Lewis, Baldassarre, Committeri, Romani, & Corbetta, 2009) which might even be correlated across multiple brain systems (Biswal, Yetkin, Haughton, & Hyde, 1995).

At the conceptual level, the brain's spontaneous activity may be compared to a concept elaborated by the philosopher Spinoza, referred to as conatus, according to which "each thing, as far as it lies in itself, strives to persevere in its being" (*Ethics*, part 3, prop. 6). In this way, the Dutch thinker outlined that living organisms have an intrinsic predisposition to survive. More recently Maturana and Varela (1991) similarly claimed that a characteristic feature of living systems is autopoiesis, that is their capacity to literally "build themselves" or to constantly self-realise fighting against entropy and the dissipation of their own energy.

The importance of such spontaneous activity for the brain, especially for conscious perception, was already mentioned in the first simulations of the GNW (Dehaene, Kerszberg, & Changeux, 1998) and shown to be necessary for conscious access in a simple cognitive task (Dumas, Nadel, Soussignan, Martinerie, & Garnero, 2010). Last in agreement with our views, the active inference theory formalizes how autonomous learning agents (whether artificial or natural) shall be endowed with a spontaneous active motivation to explore and learn (Friston et al., 2016), which other studies confirmed to be sufficient for the emergence of complex behavior without the need for immediate rewards (https://arxiv.org/abs/2205.10316).

Neural network models designed to perform continuous life-long learning have, by design, spontaneous activity that is required to learn (Lewis et al., 2009; Parisi, Kemker, Part, Kanan, & Wermter, 2019), and even the classical models of AI, namely Boltzmann machines (http://var.scholarpedia.org/article/Boltzmann_machine) have spontaneous activity at their very core which has recently fruitfully inspired neuromorphic computing (Dold et al., 2019; Petrovici, Bill, Bytschok, Schemmel, & Meier, 2016). For instance, neuromorphic research has attempted to exploit the spontaneous generation of activity, including generation of world models, and self-evaluation that are at the core of brain activities like the phases of wakefulness and sleep (Deperrois, Petrovici, Senn, & Jordan, 2022, 2024). Yet still rather unique features of spontaneous activity in humans compared to artificial systems are that: 1. this spontaneous activity is not a level of "noise" in the system (as it at present prevalently is in AI, notwithstanding attempts to achieve it without noise (Dold et al., 2019)), but well defined stochastic and globally distributed activity all over the brain, without evident sign of large scale correlations but following well-defined patterns of brain areas (Lewis et al., 2009); 2. Importance of reward and emotions in the biological brain to link the internal world to itself and to the physical, social, and cultural environment; 3. the ability to link together far-distant activities through, for instance, brain scale modulatory systems (Dehaene & Changeux, 1997, 2000), which may include cortical and non-cortical (e.g., limbic) neurons leading to the stabilization of adequate ("harmonious") relationships consciously perceived by the subject. The interconnections established between them may include epigenetically stored information unique to the life time experience of every individual subject. This means a gigantic capacity of flexibility and creativity for personal and cultural artefacts like in science (Changeux & Connes, 1995), art (Changeux, 1994), or ethics (Changeux, 2023; Evers & Changeux, 2016).

Embodiment, that is the sensorimotor abilities that humans use to explore the world through multisensory integration within a constitutive interaction between brain, body, and environment (Pennartz, 2009), plays a crucial role in humans for making possible the three features above. In fact, as a result of the interaction between brain, body, and environment humans are capable of flexible and online sensorimotor learning (i.e., of developing a realistic representation of the world that is adapted in real time) which may be intentionally (i.e., consciously) exploited. Values and the capacity for evaluation are essential for learning because without any values, or any capacity to evaluate stimuli, a system cannot learn or remember: it has to prefer some stimuli to others in order to learn. This classical idea in learning theory was expressed in neuronal terms by (Dehaene & Changeux, 1989, 1991) and by Edelman in his accounts of primary consciousness (Edelman, 1992), and was taken up in philosophical terms by (Evers, 2009) suggesting that evaluation understood as operations of selective responsiveness to reward signals is a presupposition for human consciousness to develop.

In biological organisms, embodiment is intrinsically related to the capacity to evaluate the world, discriminating between what is good and what is bad. This evaluation is mediated by emotions, reward-systems, and preferences which eventually allow the organism to discriminate between the different affordances, assigning them either positive or negative values (e.g., pleasure and pain) calibrated on the *Umwelt*. We agree with (Aru et al., 2023) that the *Umwelt* of present AI (e.g., LLMs), if any, is very limited because of its design, specifically because of how it processes information (i.e., abstract statistical processing). This impacts upon the prospect of conscious AI. Importantly, as argued by a number of cognitive scientists and philosophers, conscious processing does not depend only on the processed information, but also on internal properties of the system, as well as on its embodiment, and particularly on its emotionally charged motivations and goals, which are crucially linked to the embodiment of the agent (Damasio & Damasio, 2022; Damasio & Damasio, 2023; Roli, Jaeger, & Kauffman, 2022; Shapiro & Spaulding, 2024; Varela, Thompson, & Rosch, 2016). If so, the sole optimization of the current AI capacity for processing information is likely insufficient to achieve artificial conscious processing (Aru et al., 2023), especially if not combined with embodiment and related reward-based and emotionally charged motivations. Present AI systems may be equipped with the capacity for evaluation (Dromnelle et al., 2023), eventually instrumental for achieving the goals they have been trained for, but it is not clear if and how these goals have the same emotional correlates as in humans, so that in the end it would really feel like something for AI to achieve a goal (Boden, 2016). A relevant approach different from traditional views of AI based on outward-directed perception and abstract problem-solving has been recently proposed by (Man, Damásio, & Neven, 2022). They start from the premise that the central problem and drive of life is homeostasis (i.e., regulation of internal states instrumental to maintaining conditions compatible with life). Therefore, sense-data are meaningful if relevant to homeostatic needs. On this premise, they propose a homeostatic neural network in which a classifier has a needful and vulnerable relation with the objects it computes.

From this perspective, the limitation of LLMs is that they presently lack the multidimensional and multisensory representation of the world that in humans is mediated by the body, and therefore they are intrinsically exposed to limited and distorted knowledge eventually insufficient for a multi-modal (i.e., conscious) representation of the world. Robotics, including lately its combination with LLMs (Zhang, Chen, Li, Peng, & Mao, 2023), as well as neuromorphic AI, promise to provide AI with a form of embodiment that in principle can make possible an artificial form of multimodal experience. For example, significant results have been obtained with robotic systems endowed with the capacity for self-monitoring, which replicates at least some aspect of human self-consciousness (Chella, Frixione, & Gaglio, 2008), particularly relying on the capacity for inner speech (Chella, Pipitone, Morin, & Racy, 2020; Pipitone & Chella, 2021).

Yet the apparent lack of any emotional stake for AI in its interaction with the world remains. Given the important role for conscious experience played by the evaluative capacity of the human organism mediated by embodied features like emotions, this lack is a potential stumbling block to further investigate and to overcome on the path towards conscious AI.

In conclusion, brain features that are only partially implemented in current AI and that current research on conscious AI should try to better simulate are an embodied sensori-motor experience of the world, the spontaneous activity of the brain, which is more than simple noise as in current AI systems, autopoiesis (as the capacity for constant self-realisation), and emotions-based reward systems.

## 5. Conscious vs non-conscious processing in the brain, or res cogitans vs res extensa

The spontaneous activity referred to above has importance for the distinction between conscious and non-conscious processing, which was already suggested by Descartes via the formulation of *res cogitans* as opposed to *res extensa*, two notions that are formulated unambiguously outside a dualist context as indicated by the use of the same term "res" (Changeux & Ricoeur, 1998). In contemporary terms, it means that conscious representations "physically" – or physiologically - differ from the non-conscious ones. To distinguish between the two a masking task was introduced. When a representation progresses - for instance, in the visual pathway - from V1 to temporal cortex, a non-conscious representation is captured until a sudden divergence in activity takes place around 200 to 300 ms after stimulus onset. When the stimulus becomes conscious, there is an intense surge and propagation of additional activity, particularly when it reaches the prefrontal and parietal cortex (becomes « phenomenal »). This non-linear divergence was referred to as "ignition" (Dehaene & Changeux, 2005) and occurs regardless of the stimulus modality or paradigm used to manipulate consciousness (Del Cul, Dehaene, Reyes, Bravo, & Slachevsky, 2009; Klatzmann et al., 2023; Mashour, Roelfsema, Changeux, & Dehaene, 2020a). Spontaneous ignition may as well be recorded in the absence of external stimulation (Koukouli, Rooy, Changeux, & Maskos, 2016; Moutard, Dehaene, & Malach, 2015).

The first simulations of ignition in a global workspace architecture (Dehaene & Changeux, 2005; Dehaene, Sergent, & Changeux, 2003) already included a key role for long-distance recurrent and top-down connections in the maintenance of long-lasting sustained ignition: more stable activity would appear when sensory stimuli are consciously perceived (Schurger, Kim, & Cohen, 2015). Recently (Klatzmann et al., 2023), the hypothesis was made that the rapid onset and offset of excitatory currents mediated by AMPA receptors could help sensory areas react quickly to changes to external stimuli (Self, Kooijmans, Supèr, Lamme, & Roelfsema, 2012; Yang et al., 2018), but following the stimulus, the slow time-constant of NMDA receptors helps sustaining stimulus-specific activity in prefrontal cortex and GNW (Wang, 1999; Wang et al., 2013). This further illustrates the relevance of the biochemical diversity of the brain in conscious processing (see above).

When conscious, the "representations" show singular properties. For example, associative properties that let them organize in chains and build up forms of reasoning that do not exist non-consciously in humans, in monkeys and probably not in computers. The classical interpretation is that working memory is very shallow in monkeys (1-2 precisely 2.1 in chimps)) but far higher in humans (7+/-2). More important to our view is that conscious representations possess a singular neural organization that let them establish "physical links" into a sequence able to have a global meaning which integrate that of each component. On this basis (still largely unexplored) one should be able to build a sentence, propose a reasoning, account for language generativity (i.e., the ability to produce sentences never before said, and to understand sentences never before heard), and recursivity as the ability to place one component inside another component of the same kind and build a tree, a linguistic element or grammatical structure that can be used repeatedly in a sequence. In all these instances, again, cortical and non-cortical neurons would be involved through a large scale selection by reward systems.

LLMs have been recently claimed to have these abilities, specifically syntactic language generativity. This depends on how language generation and understanding are conceived. LLMs can generate new combinations of words on the basis of statistical patterns, and also identify the connections between those words. This kind of language production and understanding are syntactic and independent from semantics (i.e., meaning attribution possibly through selective rewards), which characterizes human language production and understanding (Bennett, 2023; Marcus & Davis, 2019). Even if LLMs are capable of fully emulating a semantically coherent human-generated language, they do so through a different strategy than human brains, which does not include meaning understanding in the usual sense (Wolfram, 2023).

The same is arguably true for reasoning and creativity. While the details of these features in the human brain are still to be clarified, data converge in indicating a central role of intrinsic and external rewards in selecting and stabilizing neuronal configuration underlying creative tasks like artistic works (Changeux, 2019). Also, the principle of degeneracy (i.e., many neuronal combinations may result in the same functional result) is the ground for the human brain plasticity and creativity, and it appears significantly limited in AI, despite relevant results emerging from neural network pruning and neuromorphic hardware.

In conclusion, brain characteristics that can help contemporary attempts to develop conscious AI systems are: a clear distinction between conscious and non-conscious representations (demarked, for instance, by the phenomenon of ignition), and the consequent singular associative properties of conscious representations; the semantic competence of the brain, which is expressed, for instance, in the capacity for symbolic understanding and mutual help; singular properties of association of conscious representations into meaningful sequences and ultimately the development of reasoning and rational thinking.

## 6. AI consciousness and social interaction challenge rational thinking and language

Starting from evolutionary and neuroscientific data, we have listed above a number of features that AI research might take into account in its attempts to advance the development of artificial consciousness. Some of the limitations of current AI that we have outlined had already been identified by other authors, such as the role of embodiment, emotions, and motivation in machine consciousness, and the issue of robot experience (Chella & Manzotti, 2009; Montemayor, 2023), while others appear less explored. Among these, the capacity for interaction between individuals in a social group which gives rise to a number of human-specific dispositions, including theory of mind, cultural experience and ethical responsibility. In principle, the fact that AI constructs have a different physical structure and a different functional organization than biological systems that are probably conscious (even if at different levels, e.g., humans vs. other species including Drosophila (Grover et al., 2022), does not imply that AI systems are not or cannot become conscious. The above analysis has revealed some factors that data indicate are correlated with conscious processing in the human brain, but this is not sufficient to conclude that they are necessary conditions for conscious processing in both humans and in other biological or in artificial systems. Notably, the issue of the actual existence of artificial forms of consciousness is importantly empirical: we need reliable indicators to identify non-biological conscious processing in AI systems (Bayne et al., 2024; L. Dung, 2023; Pennartz, Farisco, & Evers, 2019). Yet the challenge is not only empirical, but also conceptual: how can we conceptualize a non-biological kind of conscious processing? Probably the linguistic and conceptual categories that we normally use to refer to consciousness are not good (e.g., they are too anthropomorphic and anthropocentric), and should be replaced. As (Pennartz et al., 2019) outline, we may use the same term and expand our context of application accordingly, or we may coin a new term to denote machines. This choice is epistemically very important. In fact, "linguistic innovations must be well-justified in order to avoid conceptual inflation. Yet, if we are too bound by common usage traditions, we hamper development of new thought. Our languages need to evolve to enable the expression of new ideas, knowledge and normative systems." (Pennartz et al., 2019)

We do not presume to re-invent the terminology, but rather wish to recommend an improved terminological and conceptual clarity when referring consciousness to AI. This necessity may be interpreted as part of the more general call for conceptual adaptation as a consequence of the development of new technologies, as expressed in the conceptual engineering (Hopster & Löhr, 2023). An additional, specific justification of our recommendation derives from the multidimensional nature of consciousness, which makes crucial for any attempt to develop artificial consciousness to specify which is the dimension and/or the form of consciousness that it targets. Whilst the extension of other mental terms like attention and intelligence to non-human animals and AI systems may appear less problematic than consciousness, we should distinguish two semantic dimensions when referring any mental term to AI: the epistemological dimension (i.e., what is the nature of the feature indicated by the term?), and the functional dimension (i.e., how does the feature indicated by the term work? What is the relevant architecture, that is the adequate combination of structure and functionalities?). While the extension of the first dimension to AI is problematic, the extension of the second dimension may be more straightforward provided that relevant structures and functions may be replicated at the adequate scale or level of details, as argued above. In any case, a neuroscience-based linguistic and conceptual development appears plausible.

In conclusion, caution in attributing consciousness to AI, and particularly improving the conceptual clarity when making this attribution, are more than a matter of communication. It is also a matter of how thoughts develop, for language shapes thought both epistemologically, in terms of what we can think and know, and normatively, in terms of the values we develop. Therefore, these questions present us with conceptual choices that need to be made both philosophically (in terms of clarity, simplicity and logical coherence) and empirically (in terms of scientific justification, experimental validation and explanatory power).

## Conclusion

We have here reviewed some structural, evolutionary and functional features of the brain that have played an important role in making possible and/or modulating human consciousness (See Table 1). These features may possibly contribute to make artificial consciousness achievable. Against this background, we also identified some limitations of current computer hardware and AI models that we suggest should be improved for accelerating research towards the development of artificial consciousness (See Table 2).

Even if it is theoretically feasible to develop artificial systems with non-human-like forms of consciousness, we argue that taking into account the brain features above, which are presently not fully translated into AI, may accelerate and enrich the development of conscious artificial systems. This does not mean that it is actually possible to develop a human-like artificial conscious system. In fact, there is still a long way to go to fairly emulate conscious processing in humans, if it ever will be possible. Given this uncertainty, we recommend not to use for the time being the same general term (i.e., consciousness) for both humans and artificial systems; to clearly specify the key differences between them;

and, last but not least, to be very clear about which dimension, scale and level of consciousness the artificial system may possibly be capable of displaying.

**Table 1. Brain features that should be taken into account for accelerating research towards the development of conscious AI**

| | |
|---|---|
| **Organisation** | Hierarchical, nested, and multiple levels of organization, including massive biochemical and neuronal diversities (i.e., excitatory vs inhibitory neurons), connectomic relationships, circuits (i.e., thalamo-cortical loops) and a variety of bottom-up/top-down regulations between levels |
| **Rewards sensitivity** | Embodied multidimensional sensori-motor experience of the world including emotions-based reward systems and self-evaluation |
| **Physiological Activity** | Spontaneous activity contributing to conscious information processing and behavior (more than simple noise as in current AI systems) |
| | *Conatus* as striving to survive and *autopoiesis* as the capacity for constant self-realisation |
| **Evolution** | The capacity for producing variability from the same functional organization resulting in darwinian biological evolution and epigenetic development through selection and amplification |
| | Long postnatal development leading to multiple nested epigenetic synapse selections as "inprints" from the physical, biological and socio-cultural environment |
| **Degeneracy** | As a consequence of the darwinian mechanism of synase selection, different connection patterns may carry the same function, leading to plasticity, individual differences and creativity |
| **Social interaction** | Capacity for interaction between individuals in a social group which gives rise to a number of unique human-specific dispositions, including |

| | theory of mind, cultural experience and ethical responsibility |
| --- | --- |
| | Differentiation of the singularity of the individual and his/her life experience which epitomize the human person with "full conscious experience" (Ricoeur) |
| **Conscious content** | A physical (and operational) distinction between conscious and non-conscious brain representations, and the consequent development of: |
| | Semantic competence, which is expressed, for instance, by the capacity for symbolic understanding between individuals and mutual help |
| | Singular properties of association of conscious representations into meaningful sequences and ultimately the development of reasoning and rational thinking |

**Table 2. Limitations of current AI that should be ameliorated for accelerating research towards the development of conscious AI**

| |
| --- |
| Implementation of a computational approach with insufficient brain data |
| Use of an hardware made up of limited chemical elements, opposed to the highly diverse chemical composition of the brain |
| Use of an hardware which prevalently implements parallel levels of computations, opposed to the nested organization of the brain |
| Inability of current AI systems to attribute experiential value to information |
| Lack of an evolutionary and epigenetic development homologous to those of the human brain |
| Lack of an embodiment-based multidimensional and multisensory representation of the world |
| Lack of social relationships (theory of mind) and intrinsic ethical responsibility |
| Lack of differentiation of the singularity of the individual and his/her life experience which epitomize the human person with "full conscious experience" (Ricoeur) |

**Acknowledgments**
Special thanks to Jan Aru, Sacha J. van Albada, Ismael Freire, Mehdi Khamassi, and Mihai Petrovici for comments on a previous version of this paper, and to two anonymous reviewers for extremely useful comments that improved the readability and clarity of the paper.

# References

Albantakis, Barbosa, L., Findlay, G., Grasso, M., Haun, A. M., Marshall, W., . . . Tononi, G. (2023). Integrated information theory (IIT) 4.0: Formulating the properties of phenomenal existence in physical terms. *PLoS Comput Biol, 19*(10), e1011465. doi:10.1371/journal.pcbi.1011465

Alexandre, F., Dominey, P. F., Gaussier, P., Girard, B., Khamassi, M., & Rougier, N. P. (2020). When Artificial Intelligence and Computational Neuroscience Meet. In P. Marquis, O. Papini, & H. Prade (Eds.), *A Guided Tour of Artificial Intelligence Research: Volume III: Interfaces and Applications of Artificial Intelligence* (pp. 303-335). Cham: Springer International Publishing.

Ali, A., & al. (2022). Predictive coding is a consequence of energy efficiency in recurrent neural networks. *Patterns (N Y), 3*(12), 100639. doi:10.1016/j.patter.2022.100639

Arsiwalla, X. D., Solé, R., Moulin-Frier, C., Herreros, I., Sánchez-Fibla, M., & Verschure, P. (2023). The Morphospace of Consciousness: Three Kinds of Complexity for Minds and Machines. *NeuroSci, 4*(2), 79-102.

Aru, J., Larkum, M. E., & Shine, J. M. (2023). The feasibility of artificial consciousness through the lens of neuroscience. *Trends in Neurosciences, 46*(12), 1008-1017. doi:https://doi.org/10.1016/j.tins.2023.09.009

Aru, J., Suzuki, M., & Larkum, M. E. (2020). Cellular Mechanisms of Conscious Processing. *Trends Cogn Sci, 24*(10), 814-825. doi:10.1016/j.tics.2020.07.006

Barron, A. B., Halina, M., & Klein, C. (2023). Transitions in cognitive evolution. *Proc Biol Sci, 290*(2002), 20230671. doi:10.1098/rspb.2023.0671

Bartocci, M., Bergqvist, L. L., Lagercrantz, H., & Anand, K. J. (2006). Pain activates cortical areas in the preterm newborn brain. *Pain, 122*(1-2), 109-117. doi:10.1016/j.pain.2006.01.015

Bayne, T., Hohwy, J., & Owen, A. M. (2016). Are There Levels of Consciousness? *Trends Cogn Sci, 20*(6), 405-413. doi:10.1016/j.tics.2016.03.009

Bayne, T., Seth, A. K., Massimini, M., Shepherd, J., Cleeremans, A., Fleming, S. M., . . . Mudrik, L. (2024). Tests for consciousness in humans and beyond. *Trends Cogn Sci*. doi:10.1016/j.tics.2024.01.010

Bennett, M. S. (2023). *A brief history of intelligence : evolution, AI, and the five breakthroughs that made our brains* (First edition. ed.). New York: Mariner Books.

Berger, H. (1929). Über das Elektrenkephalogramm des Menschen. *Archiv für Psychiatrie und Nervenkrankheiten, 87*(1), 527-570. doi:10.1007/BF01797193

Billaudelle, S., Stradmann, Y., Schreiber, K., Cramer, B., Baumbach, A., Dold, D., . . . Meier, K. (2020, 12-14 Oct 2020). *Versatile Emulation of Spiking Neural Networks on an Accelerated Neuromorphic Substrate.* Paper presented at the 2020 IEEE International Symposium on Circuits and Systems (ISCAS).

Biswal, B., Yetkin, F. Z., Haughton, V. M., & Hyde, J. S. (1995). Functional connectivity in the motor cortex of resting human brain using echo-planar MRI. *Magn Reson Med, 34*(4), 537-541. doi:10.1002/mrm.1910340409

Block. (1995). On a confusion about a function of consciousness. *Behavioral and Brain Sciences, 18*(2), 227-287.

Block. (1997). Anti-Reductionism Slaps Back. *Noûs, 31*(s11), 107-132. doi:https://doi.org/10.1111/0029-4624.31.s11.5

Blum, L., & Blum, M. (2023). A Theoretical Computer Science Perspective on Consciousness and Artificial General Intelligence. *Engineering, 25*, 12-16. doi:https://doi.org/10.1016/j.eng.2023.03.010

Boakes, R. (1984). *From Darwin to behaviourism. Psychology and the minds of animals*. Cambridge, MA: Cambridge University Press.

Boden, M. A. (2016). *AI : its nature and future* (First edition. ed.). Oxford, United Kingdom: Oxford University Press.

Botvinick, Ritter, S., Wang, J. X., Kurth-Nelson, Z., Blundell, C., & Hassabis, D. (2019). Reinforcement Learning, Fast and Slow. *Trends Cogn Sci, 23*(5), 408-422. doi:10.1016/j.tics.2019.02.006

Botvinick, Wang, J. X., Dabney, W., Miller, K. J., & Kurth-Nelson, Z. (2020). Deep Reinforcement Learning and Its Neuroscientific Implications. *Neuron, 107*, 603-616.

Boybat, I., Le Gallo, M., Nandakumar, S. R., Moraitis, T., Parnell, T., Tuma, T., . . . Eleftheriou, E. (2018). Neuromorphic computing with multi-memristive synapses. *Nat Commun, 9*(1), 2514. doi:10.1038/s41467-018-04933-y

Buckner, R. L., & Krienen, F. M. (2013). The evolution of distributed association networks in the human brain. *Trends Cogn Sci, 17*(12), 648-665. doi:10.1016/j.tics.2013.09.017

Burke, S. M., Avstrikova, M., Noviello, C. M., Mukhtasimova, N., Changeux, J.-P., Thakur, G. A., . . . Hibbs, R. E. (2024). Structural mechanisms of α7 nicotinic receptor allosteric modulation and activation. *Cell, 187*(5), 1160-1176.e1121. doi:https://doi.org/10.1016/j.cell.2024.01.032

Butlin, P., Long, R., Elmoznino, E., Bengio, Y., Birch, J., Constant, A., . . . VanRullen, R. (2023). *Consciousness in Artificial Intelligence: Insights from the Science of Consciousness*. Retrieved from arXiv:2308.08708v3

Cao, R. (2022). Multiple realizability and the spirit of functionalism. *Synthese, 200*(6), 506. doi:10.1007/s11229-022-03524-1

Castro-Caldas, A., Petersson, K. M., Reis, A., Stone-Elander, S., & Ingvar, M. (1998). The illiterate brain. Learning to read and write during childhood influences the functional organization of the adult brain. *Brain, 121 ( Pt 6)*, 1053-1063. doi:10.1093/brain/121.6.1053

Changeux. (1986). *Neuronal man : the biology of mind*. New York: Oxford University Press.

Changeux. (1994). *Raison et plaisir*. Paris: Editions O. Jacob.

Changeux. (2006). The Ferrier Lecture 1998. The molecular biology of consciousness investigated with genetically modified mice. *Philos Trans R Soc Lond B Biol Sci, 361*(1476), 2239-2259. doi:10.1098/rstb.2006.1832

Changeux. (2017). Climbing Brain Levels of Organisation from Genes to Consciousness. *Trends Cogn Sci, 21*(3), 168-181. doi:10.1016/j.tics.2017.01.004

Changeux. (2019). 21C1Artistic Creativity: A Neuronal HypothesisSecrets of Creativity: What Neuroscience, the Arts, and Our Minds Reveal (pp. 0): Oxford University Press. Retrieved from https://doi.org/10.1093/oso/9780190462321.003.0002. doi:10.1093/oso/9780190462321.003.0002

Changeux. (2023). *Le Beau et la splendeur du vrai: Entretiens avec François L'Yvonnet*. Paris: Albin Michel.

Changeux, & Connes, A. (1995). *Conversations on mind, matter, and mathematics*. Princeton, N.J.: Princeton University Press.

Changeux, Courrège, P., & Danchin, A. (1973). A theory of the epigenesis of neuronal networks by selective stabilization of synapses. *Proc Natl Acad Sci U S A, 70*(10), 2974-2978.

Changeux, & Danchin, A. (1976). Selective stabilisation of developing synapses as a mechanism for the specification of neuronal networks. *Nature, 264*(5588), 705-712. doi:10.1038/264705a0

Changeux, Goulas, A., & Hilgetag, C. C. (2021). A Connectomic Hypothesis for the Hominization of the Brain. *Cereb Cortex, 31*(5), 2425-2449. doi:10.1093/cercor/bhaa365

Changeux, & Lou, H. C. (2011). Emergent pharmacology of conscious experience: new perspectives in substance addiction. *FASEB J, 25*(7), 2098-2108. doi:10.1096/fj.11-0702ufm

Changeux, & Ricoeur, P. (1998). *Ce Qui Nous Fait Penser. La Nature Et La Regle*. Paris: Odile Jacob.

Chella, A., Frixione, M., & Gaglio, S. (2008). A cognitive architecture for robot self-consciousness. *Artificial Intelligence in Medicine, 44*(2), 147-154. doi:https://doi.org/10.1016/j.artmed.2008.07.003

Chella, A., & Manzotti, R. (2009). MACHINE CONSCIOUSNESS: A MANIFESTO FOR ROBOTICS. *International Journal of Machine Consciousness, 01*(01), 33-51. doi:10.1142/S1793843009000062

Chella, A., Pipitone, A., Morin, A., & Racy, F. (2020). Developing Self-Awareness in Robots via Inner Speech. *Front Robot AI, 7*. doi:10.3389/frobt.2020.00016

Chirimuuta, M. (2024). *The Brain Abstracted: Simplification in the History and Philosophy of Neuroscience*: The MIT Press.

Colas, C., Karch, T., Moulin-Frier, C., & Oudeyer, P.-Y. (2022). Language and culture internalization for human-like autotelic AI. *Nature Machine Intelligence, 4*(12), 1068-1076. doi:10.1038/s42256-022-00591-4

Cramer, B., Billaudelle, S., Kanya, S., Leibfried, A., Grubl, A., Karasenko, V., . . . Zenke, F. (2022). Surrogate gradients for analog neuromorphic computing. *Proc Natl Acad Sci U S A, 119*(4). doi:10.1073/pnas.2109194119

Damasio, & Damasio, H. (2022). Homeostatic feelings and the biology of consciousness. *Brain, 145*(7), 2231-2235. doi:10.1093/brain/awac194

Damasio, & Damasio, H. (2023). Feelings Are the Source of Consciousness. *Neural Comput, 35*(3), 277-286. doi:10.1162/neco_a_01521

Dehaene-Lambertz, G., & Spelke, E. S. (2015). The Infancy of the Human Brain. *Neuron, 88*(1), 93-109. doi:10.1016/j.neuron.2015.09.026

Dehaene, S., & Changeux, J. P. (1989). A simple model of prefrontal cortex function in delayed-response tasks. *J Cogn Neurosci, 1*(3), 244-261. doi:10.1162/jocn.1989.1.3.244

Dehaene, S., & Changeux, J. P. (1991). The Wisconsin Card Sorting Test: theoretical analysis and modeling in a neuronal network. *Cereb Cortex, 1*(1), 62-79. doi:10.1093/cercor/1.1.62

Dehaene, S., & Changeux, J. P. (1997). A hierarchical neuronal network for planning behavior. *Proc Natl Acad Sci U S A, 94*(24), 13293-13298. doi:10.1073/pnas.94.24.13293

Dehaene, S., & Changeux, J. P. (2000). Reward-dependent learning in neuronal networks for planning and decision making. *Prog Brain Res, 126*, 217-229. doi:10.1016/s0079-6123(00)26016-0

Dehaene, S., & Changeux, J. P. (2005). Ongoing spontaneous activity controls access to consciousness: a neuronal model for inattentional blindness. *PLoS Biol, 3*(5), e141. doi:10.1371/journal.pbio.0030141

Dehaene, S., & Changeux, J. P. (2011). Experimental and theoretical approaches to conscious processing. *Neuron, 70*(2), 200-227. doi:10.1016/j.neuron.2011.03.018

Dehaene, S., Kerszberg, M., & Changeux, J. P. (1998). A neuronal model of a global workspace in effortful cognitive tasks. *Proc Natl Acad Sci U S A, 95*(24), 14529-14534.

Dehaene, S., Lau, H., & Kouider, S. (2017). What is consciousness, and could machines have it? *Science, 358*(6362), 486-492. doi:10.1126/science.aan8871

Dehaene, S., Pegado, F., Braga, L. W., Ventura, P., Nunes Filho, G., Jobert, A., . . . Cohen, L. (2010). How learning to read changes the cortical networks for vision and language. *Science, 330*(6009), 1359-1364. doi:10.1126/science.1194140

Dehaene, S., Sergent, C., & Changeux, J. P. (2003). A neuronal network model linking subjective reports and objective physiological data during conscious perception. *Proc Natl Acad Sci U S A, 100*(14), 8520-8525. doi:10.1073/pnas.1332574100

Del Cul, A., Dehaene, S., Reyes, P., Bravo, E., & Slachevsky, A. (2009). Causal role of prefrontal cortex in the threshold for access to consciousness. *Brain, 132*(Pt 9), 2531-2540. doi:10.1093/brain/awp111

Deperrois, N., Petrovici, M. A., Senn, W., & Jordan, J. (2022). Learning cortical representations through perturbed and adversarial dreaming. *eLife, 11*. doi:10.7554/eLife.76384

Deperrois, N., Petrovici, M. A., Senn, W., & Jordan, J. (2024). Learning beyond sensations: How dreams organize neuronal representations. *Neurosci Biobehav Rev, 157*, 105508. doi:10.1016/j.neubiorev.2023.105508

Dold, D., Bytschok, I., Kungl, A. F., Baumbach, A., Breitwieser, O., Senn, W., . . . Petrovici, M. A. (2019). Stochasticity from function — Why the Bayesian brain may need no noise. *Neural Networks, 119*, 200-213. doi:https://doi.org/10.1016/j.neunet.2019.08.002

Dromnelle, R., Renaudo, E., Chetouani, M., Maragos, P., Chatila, R., Girard, B., & Khamassi, M. (2023). Reducing Computational Cost During Robot Navigation and Human–Robot Interaction with a Human-Inspired Reinforcement Learning Architecture. *International Journal of Social Robotics, 15*(8), 1297-1323. doi:10.1007/s12369-022-00942-6

Dubois, J., Kostovic, I., & Judas, M. (2015). Development of structural and functional connectivity.

Dumas, G., Nadel, J., Soussignan, R., Martinerie, J., & Garnero, L. (2010). Inter-brain synchronization during social interaction. *PLoS One, 5*(8), e12166. doi:10.1371/journal.pone.0012166

Dung, & Newen, A. (2023). Profiles of animal consciousness: A species-sensitive, two-tier account to quality and distribution. *Cognition, 235*, 105409. doi:10.1016/j.cognition.2023.105409

Dung, L. (2023). Tests of Animal Consciousness are Tests of Machine Consciousness. *Erkenntnis*. doi:10.1007/s10670-023-00753-9

Edelman. (1992). *Bright air, brilliant fire : on the matter of the mind*. New York, NY: BasicBooks.

Edelman, & Gally, J. A. (2001). Degeneracy and complexity in biological systems. *Proc Natl Acad Sci U S A, 98*(24), 13763-13768. doi:10.1073/pnas.231499798

Edelman, & Mountcastle, V. B. (1978). *The mindful brain: Cortical organization and the group-selective theory of higher brain function*. Oxford, England: Massachusetts Inst of Technology Pr.

Eliasmith, C., Stewart, T. C., Choo, X., Bekolay, T., DeWolf, T., Tang, Y., & Rasmussen, D. (2012). A large-scale model of the functioning brain. *Science, 338*(6111), 1202-1205. doi:10.1126/science.1225266

Esser, S. K., Merolla, P. A., Arthur, J. V., Cassidy, A. S., Appuswamy, R., Andreopoulos, A., . . . Modha, D. S. (2016). Convolutional networks for fast, energy-efficient neuromorphic computing. *Proc Natl Acad Sci U S A, 113*(41), 11441-11446. doi:10.1073/pnas.1604850113

Evers. (2009). *Neuroetique. Quand la matière s'éveille*. Paris: Odile Jacob.

Evers. (2015). Can we be epigenetically proactive? In W. Metzinger T., J. (Ed.), *Open Mind: Philosophy and the mind sciences in the 21st century*. Cambridge, MA: MIT Press.

Evers, & Changeux, J. P. (2016). Proactive epigenesis and ethical innovation: A neuronal hypothesis for the genesis of ethical rules. *EMBO Rep, 17*(10), 1361-1364. doi:10.15252/embr.201642783

Evers, & Sigman, M. (2013). Possibilities and limits of mind-reading: A neurophilosophical perspective. *Consciousness and Cognition, 22*, 887-897.

Farisco, M. (2024). The ethical implications of indicators of consciousness in artificial systems *Developments in Neuroethics and Bioethics*: Academic Press.

Farisco, M., Baldassarre, G., Cartoni, E., Leach, A., Petrovici, M. A., Rosemann, A., . . . Van Albada, S. J. (2023). A method for the ethical analysis of brain-inspired AI. Retrieved from arXiv:2305.10938v1 website:

Farisco, M., Kotaleski, J. H., & Evers, K. (2018). Large-Scale Brain Simulation and Disorders of Consciousness. Mapping Technical and Conceptual Issues. *Front Psychol, 9*, 585. doi:10.3389/fpsyg.2018.00585

Farisco, M., Laureys, S., & Evers, K. (2015). Externalization of consciousness. Scientific possibilities and clinical implications. *Curr Top Behav Neurosci, 19*, 205-222. doi:10.1007/7854_2014_338

Floreano, D., Ijspeert, A. J., & Schaal, S. (2014). Robotics and neuroscience. *Curr Biol, 24*(18), R910-R920. doi:10.1016/j.cub.2014.07.058

Floreano, D., & Mattiussi, C. (2008). *Bio-Inspired Artificial Intelligence*. Cambridge, MA: MIT Press.

Friston, K., FitzGerald, T., Rigoli, F., Schwartenbeck, P., O□Doherty, J., & Pezzulo, G. (2016). Active inference and learning. *Neuroscience & Biobehavioral Reviews, 68*, 862-879. doi:https://doi.org/10.1016/j.neubiorev.2016.06.022

Godfrey-Smith, P. (2023). *Nervous Systems, Functionalism, and Artificial Minds*. Retrieved from https://petergodfreysmith.com/wp-content/uploads/2023/12/NYU-Oct-2023-Animals-AI-Functionalism-paper-Post-C3.pdf

Göltz, J., Kriener, L., Sabado, V., & Petrovici, M. A. (2021). Fast and Energy-efficient Deep Neuromorphic Learning. *ERCIM NEWS, 125*, 17-18.

Grillner, S., Deliagina, T., Ekeberg, O., el Manira, A., Hill, R. H., Lansner, A., . . . Wallén, P. (1995). Neural networks that co-ordinate locomotion and body orientation in lamprey. *Trends Neurosci, 18*(6), 270-279.

Grondin, S. (2001). From physical time to the first and second moments of psychological time. *Psychol Bull, 127*(1), 22-44. doi:10.1037/0033-2909.127.1.22

Grover, D., Chen, J. Y., Xie, J., Li, J., Changeux, J. P., & Greenspan, R. J. (2022). Differential mechanisms underlie trace and delay conditioning in Drosophila. *Nature, 603*(7900), 302-308. doi:10.1038/s41586-022-04433-6

Guerguiev, J., Lillicrap, T. P., & Richards, B. A. (2017). Towards deep learning with segregated dendrites. *eLife, 6*. doi:10.7554/eLife.22901

Gupta, A., Savarese, S., Ganguli, S., & Fei-Fei, L. (2021). Embodied intelligence via learning and evolution. *Nature Communications, 12*(1), 5721. doi:10.1038/s41467-021-25874-z

Haider, P., Ellenberger, B., Kriener, L., Jordan, J., Senn, W., & Petrovici, M. A. (2021). Latent equilibrium: A unified learning theory for arbitrarily fast computation with arbitrarily slow neurons. *Advances in Neural Information Processing Systems, 34*, 17839-17851.

Hassabis, D., Kumaran, D., Summerfield, C., & Botvinick, M. (2017). Neuroscience-Inspired Artificial Intelligence. *Neuron, 95*(2), 245-258. doi:10.1016/j.neuron.2017.06.011

Hildt, E. (2023). The Prospects of Artificial Consciousness: Ethical Dimensions and Concerns. *AJOB Neurosci, 14*(2), 58-71. doi:10.1080/21507740.2022.2148773

Hoel, E. P. (2017). When the Map Is Better Than the Territory. *Entropy, 19*(5), 188.

Hopster, J., & Löhr, G. (2023). Conceptual Engineering and Philosophy of Technology: Amelioration or Adaptation? *Philosophy & Technology, 36*(4), 70. doi:10.1007/s13347-023-00670-3

Hublin, J.-J., & Changeux, J. P. (2022). Paleoanthropology of cognition: an overview on Hominins brain evolution. *Comptes Rendus Biologies, 345*(2), 57-75.

Humphries, M. D., Khamassi, M., & Gurney, K. (2012). Dopaminergic Control of the Exploration-Exploitation Trade-Off via the Basal Ganglia. *Front Neurosci, 6*, 9. doi:10.3389/fnins.2012.00009

Irwin, L. N. (2024). Behavioral indicators of heterogeneous subjective experience in animals across the phylogenetic spectrum: Implications for comparative animal phenomenology. *Heliyon*. doi:10.1016/j.heliyon.2024.e28421

Jékely, G. (2021). The chemical brain hypothesis for the origin of nervous systems. *Philos Trans R Soc Lond B Biol Sci, 376*(1821), 20190761. doi:10.1098/rstb.2019.0761

Jordan, J., Schmidt, M., Senn, W., & Petrovici, M. A. (2021). Evolving interpretable plasticity for spiking networks. *eLife, 10*, e66273. doi:10.7554/eLife.66273

Kanaev, I. A. (2022). Evolutionary origin and the development of consciousness. *Neuroscience & Biobehavioral Reviews, 133*, 104511. doi:https://doi.org/10.1016/j.neubiorev.2021.12.034

Kasthuri, N., & Lichtman, J. W. (2003). The role of neuronal identity in synaptic competition. *Nature, 424*(6947), 426-430. doi:10.1038/nature01836

Kelty-Stephen, D., Cisek, P. E., De Bari, B., Dixon, J., Favela, L. H., Hasselman, F., . . . Mangalam, M. (2022). In search for an alternative to the computer metaphor of the mind and brain. Retrieved from arXiv:2206.04603

Klatzmann, U., Froudist-Walsh, S., Bliss, D., Theodoni, P., Mejías, J., Niu, M., . . . Wang, X.-J. (2023). A connectome-based model of conscious access in monkey cortex. *bioRxiv*, 2022.2002.2020.481230. doi:10.1101/2022.02.20.481230

Kleiner, J. (2024). *Consciousness qua Mortal Computation*.

Kouider, S., Stahlhut, C., Gelskov, S. V., Barbosa, L. S., Dutat, M., de Gardelle, V., . . . Dehaene-Lambertz, G. (2013). A neural marker of perceptual consciousness in infants. *Science, 340*(6130), 376-380. doi:10.1126/science.1232509

Koukouli, F., Rooy, M., Changeux, J. P., & Maskos, U. (2016). Nicotinic receptors in mouse prefrontal cortex modulate ultraslow fluctuations related to conscious processing. *Proc Natl Acad Sci U S A, 113*(51), 14823-14828. doi:10.1073/pnas.1614417113

Lagercrantz. (2016). *Infant Brain Development : Formation of the Mind and the Emergence of Consciousness* (pp. 1 online resource (XI, 156 pages 195 illustrations, 170 illustrations in color). doi:10.1007/978-3-319-44845-9

Lagercrantz, & Changeux, J. P. (2009). The emergence of human consciousness: from fetal to neonatal life. *Pediatr Res, 65*(3), 255-260. doi:10.1203/PDR.0b013e3181973b0d

Lagercrantz, & Changeux, J. P. (2010). Basic consciousness of the newborn. *Semin Perinatol, 34*(3), 201-206. doi:10.1053/j.semperi.2010.02.004

LeCun, Y., Bengio, Y., & Hinton, G. (2015). Deep learning. *Nature, 521*(7553), 436-444. doi:10.1038/nature14539

LeDoux, Birch, J., Andrews, K., Clayton, N. S., Daw, N. D., Frith, C., . . . Vandekerckhove, M. M. P. (2023). Consciousness beyond the human case. *Current Biology, 33*(16), R832-R840. doi:https://doi.org/10.1016/j.cub.2023.06.067

Lenharo, M. (2024). AI consciousness: scientists say we urgently need answers. *Nature, 625*(7994), 226. doi:10.1038/d41586-023-04047-6

Levine, J. (1983). Materialism and qualia: the explanatory gap. *Pacific Philosophical Quarterly, 64*, 354-361.

Lewis, C. M., Baldassarre, A., Committeri, G., Romani, G. L., & Corbetta, M. (2009). Learning sculpts the spontaneous activity of the resting human brain. *Proceedings of the National Academy of Sciences, 106*(41), 17558-17563. doi:doi:10.1073/pnas.0902455106

Lillicrap, T. P., Santoro, A., Marris, L., Akerman, C. J., & Hinton, G. (2020). Backpropagation and the brain. *Nat Rev Neurosci, 21*(6), 335-346. doi:10.1038/s41583-020-0277-3

Lou, H. C., Changeux, J. P., & Rosenstand, A. (2016). Towards a cognitive neuroscience of self-awareness. *Neurosci Biobehav Rev*. doi:10.1016/j.neubiorev.2016.04.004

Lou, H. C., Changeux, J. P., & Rosenstand, A. (2017). Towards a cognitive neuroscience of self-awareness. *Neurosci Biobehav Rev, 83*, 765-773. doi:10.1016/j.neubiorev.2016.04.004

Man, K., Damásio, A. S., & Neven, H. (2022). Need is All You Need: Homeostatic Neural Networks Adapt to Concept Shift. *ArXiv, abs/2205.08645*.

Marcus, G., & Davis, E. (2019). *Rebooting AI : building artificial intelligence we can trust* (First edition. ed.). New York: Pantheon Books.

Mashour, Roelfsema, Changeux, & Dehaene. (2020a). Conscious Processing and the Global Neuronal Workspace Hypothesis. *Neuron, 105*(5), 776-798. doi:10.1016/j.neuron.2020.01.026

Mashour, Roelfsema, P., Changeux, J.-P., & Dehaene, S. (2020b). Conscious processing and the global neuronal workspace hypothesis. *Neuron, 105*(5), 776-798.

Max, K., Kriener, L., Pineda García, G., Nowotny, T., Senn, W., & Petrovici, M. A. (2023, 2023//). *Learning Efficient Backprojections Across Cortical Hierarchies in Real Time.* Paper presented at the Artificial Neural Networks and Machine Learning – ICANN 2023, Cham.

McCulloch, W., & Pitts, W. (1943). A Logical Calculus of Ideas Immanent in Nervous Activity. *Bulletin of Mathematical Biophysics, 4*(4), 115-133.

Melis, A. P., & Raihani, N. J. (2023). The cognitive challenges of cooperation in human and nonhuman animals. *Nature Reviews Psychology, 2*(9), 523-536. doi:10.1038/s44159-023-00207-7

Mesoudi, A., Laland, K. N., Boyd, R., Buchanan, B., Flynn, E., McCauley, R. N., . . . Tennie, C. (2013). 193The Cultural Evolution of Technology and Science. In P. J. Richerson & M. H. Christiansen (Eds.), *Cultural Evolution: Society, Technology, Language, and Religion* (pp. 0): The MIT Press.

Metzinger, T. (2021). An Argument for a Global Moratorium onSynthetic Phenomenology. *Journal of Arti˜cial Intelligence and Consciousness, 8*(1), 1-24.

Miglino, O., Lund, H. H., & Nolfi, S. (1995). Evolving mobile robots in simulated and real environments. *Artif Life, 2*(4), 417-434. doi:10.1162/artl.1995.2.4.417

Minsky, M., & Papert, S. A. (2017). *Perceptrons: An Introduction to Computational Geometry*: The MIT Press.

Mitchell, M. (2023). How do we know how smart AI systems are? *Science, 381*(6654), adj5957. doi:10.1126/science.adj5957

Mitchell, M., & Krakauer, D. C. (2023). The debate over understanding in AI's large language models. *Proc Natl Acad Sci U S A, 120*(13), e2215907120. doi:10.1073/pnas.2215907120

Momennejad, I. (2023). A rubric for human-like agents and NeuroAI. *Philosophical Transactions of the Royal Society B: Biological Sciences, 378*(1869), 20210446. doi:doi:10.1098/rstb.2021.0446

Montemayor, C. (2023). *The Prospect of a Humanitarian Artificial Intelligence : agency and value alignment*.

Moulin-Frier, C., Arsiwalla, X. D., Puigbo, J.-Y., Sánchez-Fibla, M., Duff, A., & Verschure, P. F. M. J. (2016). *Top-Down and Bottom-Up Interactions between Low-Level Reactive Control and Symbolic Rule Learning in Embodied Agents*. https://ceur-ws.org/Vol-1773/CoCoNIPS_2016_paper8.pdf

Moutard, C., Dehaene, S., & Malach, R. (2015). Spontaneous Fluctuations and Non-linear Ignitions: Two Dynamic Faces of Cortical Recurrent Loops. *Neuron, 88*(1), 194-206. doi:10.1016/j.neuron.2015.09.018

Nolfi, S., Parisi, D., & Elman, J. L. (1994). Learning and Evolution in Neural Networks. *Adaptive Behavior, 3*(1), 5-28. doi:10.1177/105971239400300102

Oliveira, A. L. (2022). A blueprint for conscious machines. *Proceedings of the National Academy of Sciences, 119*(23), e2205971119. doi:10.1073/pnas.2205971119

Parisi, G. I., Kemker, R., Part, J. L., Kanan, C., & Wermter, S. (2019). Continual lifelong learning with neural networks: A review. *Neural Networks, 113*, 54-71. doi:https://doi.org/10.1016/j.neunet.2019.01.012

Park, Lee, J., & Jeon, D. (2019, 17-21 Feb. 2019). *7.6 A 65nm 236.5nJ/Classification Neuromorphic Processor with 7.5% Energy Overhead On-Chip Learning Using Direct Spike-Only Feedback.* Paper presented at the 2019 IEEE International Solid- State Circuits Conference - (ISSCC).

Pennartz. (2009). Identification and integration of sensory modalities: neural basis and relation to consciousness. *Conscious Cogn, 18*(3), 718-739. doi:10.1016/j.concog.2009.03.003

Pennartz. (2024). *The consciousness network : how the brain creates our reality*. Abingdon, Oxon ; New York, NY: Routledge.

Pennartz, Farisco, M., & Evers, K. (2019). Indicators and Criteria of Consciousness in Animals and Intelligent Machines: An Inside-Out Approach. *Front Syst Neurosci, 13*, 25. doi:10.3389/fnsys.2019.00025

Petrovici, Bill, J., Bytschok, I., Schemmel, J., & Meier, K. (2016). Stochastic inference with spiking neurons in the high-conductance state. *Physical Review E, 94*(4), 042312. doi:10.1103/PhysRevE.94.042312

Petrovici, Vogginger, B., Muller, P., Breitwieser, O., Lundqvist, M., Muller, L., . . . Meier, K. (2014). Characterization and compensation of network-level anomalies in mixed-signal neuromorphic modeling platforms. *PLoS One, 9*(10), e108590. doi:10.1371/journal.pone.0108590

Pezzulo, G., Parr, T., Cisek, P., Clark, A., & Friston, K. (2024). Generating meaning: active inference and the scope and limits of passive AI. *Trends Cogn Sci, 28*(2), 97-112. doi:10.1016/j.tics.2023.10.002

Pfeil, T., Grübl, A., Jeltsch, S., Müller, E., Müller, P., Petrovici, M. A., . . . Meier, K. (2013). Six networks on a universal neuromorphic computing substrate. *Front Neurosci, 7*, 11. doi:10.3389/fnins.2013.00011

Phillips, W. A. (2023). *The cooperative neuron*. New York: Oxford University Press.

Piccinini, G. (2022). Situated Neural Representations: Solving the Problems of Content. *Front Neurorobot, 16*, 846979. doi:10.3389/fnbot.2022.846979

Pipitone, A., & Chella, A. (2021). Robot passes the mirror test by inner speech. *Robotics and Autonomous Systems, 144*, 103838. doi:https://doi.org/10.1016/j.robot.2021.103838

Poo, M.-m. (2018). Towards brain-inspired artificial intelligence. *National Science Review, 5*(6), 785-785. doi:10.1093/nsr/nwy120

Posner, M. I., & Rothbart, M. K. (2007). *Educating the human brain*. Washington, DC, US: American Psychological Association.

Raiteri, M. (2006). Functional pharmacology in human brain. *Pharmacol Rev, 58*(2), 162-193. doi:10.1124/pr.58.2.5

Rochat, P. (2003). Five levels of self-awareness as they unfold early in life. *Consciousness and Cognition: An International Journal, 12*(4), 717-731. doi:10.1016/S1053-8100(03)00081-3

Roli, Jaeger, J., & Kauffman, S. A. (2022). How Organisms Come to Know the World: Fundamental Limits on Artificial General Intelligence. *Frontiers in Ecology and Evolution, 9*. doi:10.3389/fevo.2021.806283

Rosas, F. E., Mediano, P. A. M., Jensen, H. J., Seth, A. K., Barrett, A. B., Carhart-Harris, R. L., & Bor, D. (2020). Reconciling emergences: An information-theoretic approach to identify causal emergence in multivariate data. *PLoS Comput Biol, 16*(12), e1008289. doi:10.1371/journal.pcbi.1008289

Sandved-Smith, L., Hesp, C., Mattout, J., Friston, K., Lutz, A., & Ramstead, M. J. D. (2021). Towards a computational phenomenology of mental action: modelling meta-awareness and attentional control with deep parametric active inference. *Neuroscience of Consciousness, 2021*(1). doi:10.1093/nc/niab018

Scellier, B., & Bengio, Y. (2017). Equilibrium Propagation: Bridging the Gap between Energy-Based Models and Backpropagation. *Front Comput Neurosci, 11*, 24. doi:10.3389/fncom.2017.00024

Schurger, A., Kim, M.-S., & Cohen, J. D. (2015). Paradoxical interaction between ocular activity, perception, and decision confidence at the threshold of vision. *PLoS One, 10*(5). doi:10.1371/journal.pone.0125278

Searle, J. R. (2000). Consciousness. *Annu Rev Neurosci, 23*, 557-578. doi:10.1146/annurev.neuro.23.1.557

Self, M. W., Kooijmans, R. N., Supèr, H., Lamme, V. A., & Roelfsema, P. R. (2012). Different glutamate receptors convey feedforward and recurrent processing in macaque V1. *Proc Natl Acad Sci U S A, 109*(27), 11031-11036. doi:10.1073/pnas.1119527109

Senn, W., Dold, D., Kungl, A. F., Ellenberger, B., Jordan, J., Bengio, Y., . . . Petrovici, M. A. (2023). A Neuronal Least-Action Principle for Real-Time Learning in Cortical Circuits. *bioRxiv*, 2023.2003.2025.534198. doi:10.1101/2023.03.25.534198

Seth. (2021). *Being you the inside story of your inner universe* (pp. 1 online resource). Retrieved from http://link.overdrive.com/?websiteID=110056&titleID=5068666

Seth. (2024). Conscious artificial intelligence and biological naturalism. Retrieved from https://doi.org/10.31234/osf.io/tz6an website:

Shanahan. (2024). Talking about Large Language Models. *Commun. ACM, 67*(2), 68–79. doi:10.1145/3624724

Shanahan, Crosby, M., Beyret, B., & Cheke, L. (2021). Artificial Intelligence and the Common Sense of Animals: (Trends in Cognitive Sciences 24, 862-872, 2020). *Trends Cogn Sci, 25*(2), 172. doi:10.1016/j.tics.2020.10.008

Shapiro, L. A., & Spaulding, S. (2024). *The Routledge handbook of embodied cognition* (Second edition. ed.). Abingdon, Oxon ; New York, NY: Routledge.

Silver, D., Hubert, T., Schrittwieser, J., Antonoglou, I., Lai, M., Guez, A., . . . Hassabis, D. (2018). A general reinforcement learning algorithm that masters chess, shogi, and Go through self-play. *Science, 362*(6419), 1140-1144. doi:10.1126/science.aar6404

Smaldino, P. E. (2017). *Models Are Stupid, and We Need More of Them*.

Solée, R. V., Valverde, S., Casals, M. R., Kauffman, S. A., Farmer, D., & Eldredge, N. (2013). The evolutionary ecology of technological innovations. *Complexity, 18*(4), 15-27. doi:https://doi.org/10.1002/cplx.21436

Stent, G. S., Kristan, W. B., Jr., Friesen, W. O., Ort, C. A., Poon, M., & Calabrese, R. L. (1978). Neuronal generation of the leech swimming movement. *Science, 200*(4348), 1348-1357. doi:10.1126/science.663615

Thompson, E. (2018). Biopsychism, Minimal Life, and Sentience.   Retrieved from https://psa2018.philsci.org/user-profile/abstract/public/352/biopsychism-minimal-life-and-sentience

Tomasello, M. (2022). *The evolution of agency : behavioral organization from lizards to humans*. Cambridge, Massachusetts: The MIT Press.

Tononi. (2015). Integrated information theory *Scholapedia* (Vol. 10 (1)).

Tononi, & Edelman, G. M. (1998). Consciousness and complexity. *Science, 282*(5395), 1846-1851. doi:10.1126/science.282.5395.1846

Tononi, G., Sporns, O., & Edelman, G. M. (1999). Measures of degeneracy and redundancy in biological networks. *Proceedings of the National Academy of Sciences, 96*(6), 3257-3262. doi:doi:10.1073/pnas.96.6.3257

Ugur, B., Chen, K., & Bellen, H. J. (2016). Drosophila tools and assays for the study of human diseases. *Dis Model Mech, 9*(3), 235-244. doi:10.1242/dmm.023762

Valverde, S. (2016). Major transitions in information technology. *Philos Trans R Soc Lond B Biol Sci, 371*(1701). doi:10.1098/rstb.2015.0450

van Rooij, I., Guest, O., Adolfi, F., de Haan, R., Kolokolova, A., & Rich, P. (2023). Reclaiming AI as a theoretical tool for cognitive science. Retrieved from https://osf.io/preprints/psyarxiv/4cbuv website:

VanRullen, R., & Kanai, R. (2021). Deep learning and the Global Workspace Theory. *Trends in Neurosciences, 44*(9), 692-704. doi:https://doi.org/10.1016/j.tins.2021.04.005

Varela, F. J., Thompson, E., & Rosch, E. (2016). *The embodied mind : cognitive science and human experience* (revised edition. ed.). Cambridge, Massachusetts ; London England: MIT Press.

Verschure, P. F. (2016). Synthetic consciousness: the distributed adaptive control perspective. *Philos Trans R Soc Lond B Biol Sci, 371*(1701). doi:10.1098/rstb.2015.0448

Vinyals, O., Babuschkin, I., Czarnecki, W. M., Mathieu, M., Dudzik, A., Chung, J., . . . Silver, D. (2019). Grandmaster level in StarCraft II using multi-agent reinforcement learning. *Nature, 575*(7782), 350-354. doi:10.1038/s41586-019-1724-z

Volzhenin, K., Changeux, J. P., & Dumas, G. (2022). Multilevel development of cognitive abilities in an artificial neural network. *Proc Natl Acad Sci U S A, 119*(39), e2201304119. doi:10.1073/pnas.2201304119

Waldrop, M. M. (2019). What are the limits of deep learning? *Proceedings of the National Academy of Sciences, 116*(4), 1074-1077. doi:doi:10.1073/pnas.1821594116

Walter, J. (2021). Consciousness as a multidimensional phenomenon: implications for the assessment of disorders of consciousness. *Neurosci Conscious, 2021*(2), niab047. doi:10.1093/nc/niab047

Wang. (1999). Synaptic basis of cortical persistent activity: the importance of NMDA receptors to working memory. *J Neurosci, 19*(21), 9587-9603. doi:10.1523/jneurosci.19-21-09587.1999

Wang, Agrawal, A., Yu, E., & Roy, K. (2021). Multi-Level Neuromorphic Devices Built on Emerging Ferroic Materials: A Review. *Front Neurosci, 15*, 661667. doi:10.3389/fnins.2021.661667

Wang, Yang, Y., Wang, C. J., Gamo, N. J., Jin, L. E., Mazer, J. A., . . . Arnsten, A. F. (2013). NMDA receptors subserve persistent neuronal firing during working memory in dorsolateral prefrontal cortex. *Neuron, 77*(4), 736-749. doi:10.1016/j.neuron.2012.12.032

Whittington, J. C. R., & Bogacz, R. (2017). An Approximation of the Error Backpropagation Algorithm in a Predictive Coding Network with Local Hebbian Synaptic Plasticity. *Neural Comput, 29*(5), 1229-1262. doi:10.1162/NECO_a_00949

Wingo, A. P., Dammer, E. B., Breen, M. S., Logsdon, B. A., Duong, D. M., Troncosco, J. C., . . . Wingo, T. S. (2019). Large-scale proteomic analysis of human brain identifies proteins associated with cognitive trajectory in advanced age. *Nat Commun, 10*(1), 1619. doi:10.1038/s41467-019-09613-z

Wolfram, S. (2023). What Is ChatGPT Doing ... and Why Does It Work? Retrieved from writings.stephenwolfram.com/2023/02/what-is-chatgpt-doing-and-why-does-it-work website:

Yang, C., Kobayashi, S., Nakao, K., Dong, C., Han, M., Qu, Y., . . . Hashimoto, K. (2018). AMPA Receptor Activation-Independent Antidepressant Actions of Ketamine Metabolite (S)-Norketamine. *Biol Psychiatry, 84*(8), 591-600. doi:10.1016/j.biopsych.2018.05.007

Zador, A., Escola, S., Richards, B., Olveczky, B., Bengio, Y., Boahen, K., . . . Tsao, D. (2023). Catalyzing next-generation Artificial Intelligence through NeuroAI. *Nat Commun, 14*(1), 1597. doi:10.1038/s41467-023-37180-x

Zelazo, P. D., Craik, F. I., & Booth, L. (2004). Executive function across the life span. *Acta Psychol (Amst), 115*(2-3), 167-183. doi:10.1016/j.actpsy.2003.12.005

Zhang, C., Chen, J., Li, J., Peng, Y., & Mao, Z. (2023). Large language models for human–robot interaction: A review. *Biomimetic Intelligence and Robotics, 3*(4), 100131. doi:https://doi.org/10.1016/j.birob.2023.100131